\documentclass[11pt]{article}
\usepackage{amsmath,amssymb,color,cite}

\usepackage{amsfonts}

\def\tc{{{\hat c}}}
\def\ts{{{\hat s}}}
\def\cA{{{\cal A}}}

\def\sA{{{\sst A}}}
\def\sB{{{\sst B}}}
\def\sC{{{\sst C}}}
\def\sD{{{\sst D}}}

\def\oneone{\rlap 1\mkern4mu{\rm l}}

\newcommand{\hoch}[1]{$\, ^{#1}$}

\makeatletter
\@addtoreset{equation}{section}
\makeatother

\newcommand{\be}{\begin{equation}}
\newcommand{\ee}{\end{equation}}
\newcommand{\bea}{\setlength\arraycolsep{2pt} \begin{eqnarray}}
\newcommand{\eea}{\end{eqnarray}}
\newcommand{\nn}{\nonumber}

\def\ft#1#2{{\textstyle{\frac{\scriptstyle #1}{\scriptstyle #2} } }}
\def\fft#1#2{{\frac{#1}{#2}}}

\def\0{{\sst{(0)}}}
\def\1{{\sst{(1)}}}
\def\2{{\sst{(2)}}}
\def\3{{\sst{(3)}}}
\def\4{{\sst{(4)}}}
\def\5{{\sst{(5)}}}
\def\6{{\sst{(6)}}}
\def\7{{\sst{(7)}}}
\def\8{{\sst{(8)}}}
\def\sst#1{{\scriptscriptstyle #1}}
\def\oneone{\rlap 1\mkern4mu{\rm l}}
\def\ep{{\epsilon}}
\def\del{{\partial}}

\def\im{{{\rm i}}}

\def\supercramp{\medmuskip = 0.1mu plus 1mu minus 1mu}

\def\uncramp{\medmuskip = 4mu plus 2mu minus 4mu}

\begin{document}

\begin{flushright}
\hfill{ \
MIFPA-14-04\ \ \ \ }
\end{flushright}

\vspace{25pt}
\begin{center}
{\Large {\bf An $\omega$ Deformation of Gauged STU Supergravity}
}

\vspace{30pt}

{\Large
H. L\"u\hoch{1}, Yi Pang\hoch{2} and C.N. Pope\hoch{2,3}
}

\vspace{10pt}

\hoch{1}{\it Department of Physics, Beijing Normal University,
Beijing 100875, China}

\vspace{10pt}

\hoch{2} {\it George P. \& Cynthia Woods Mitchell  Institute
for Fundamental Physics and Astronomy,\\
Texas A\&M University, College Station, TX 77843, USA}

\vspace{10pt}

\hoch{3}{\it DAMTP, Centre for Mathematical Sciences,
 Cambridge University,\\  Wilberforce Road, Cambridge CB3 OWA, UK}

\vspace{20pt}

\underline{ABSTRACT}
\end{center}
\vspace{15pt}

   Four-dimensional ${\cal N}=2$ gauged STU supergravity is
a consistent truncation of the standard
${\cal N}=8$ gauged $SO(8)$ supergravity in
which just the four $U(1)$ gauge fields in the Cartan subgroup of
$SO(8)$ are retained.  One of these is the graviphoton in the ${\cal N}=2$
supergravity multiplet and the other three lie in three vector multiplets.
In this paper we carry out the analogous consistent truncation of the
newly-discovered family of $\omega$-deformed ${\cal N}=8$ gauged $SO(8)$
supergravities, thereby obtaining a family of $\omega$-deformed
STU gauged supergravities.  Unlike in some other truncations of the
deformed ${\cal N}=8$ supergravity that have been considered, here the
scalar potential of the deformed STU theory is independent of the
$\omega$ parameter.  However, it enters in the scalar
couplings in the gauge-field kinetic terms, and it is
non-trivial because of the minimal couplings of the fermion fields
to the gauge potentials.  We discuss the supersymmetry transformation
rules in the $\omega$-deformed supergravities, and present some examples
of black hole solutions.

\thispagestyle{empty}

\pagebreak
\voffset=-40pt
\setcounter{page}{1}

\tableofcontents

\addtocontents{toc}{\protect\setcounter{tocdepth}{2}}



\section{Introduction}

   It had long been thought that the maximal ${\cal N}=8$
gauged $SO(8)$ supergravity theory that was constructed by
de Wit and Nicolai in 1982 \cite{dewitnic} was unique.
However, based on a gauging-independent formulation of the theory
\cite{deWit:2007mt},
evidence was recently found indicating that in fact there
exists a one-parameter
family of inequivalent ${\cal N}=8$
gauged $SO(8)$ supergravity theories, characterised by an angular
parameter $\omega$ \cite{dalinvtri}.  When written in the purely
electric frame,
the new theories were given in \cite{dewitnicnew}.

   There have subsequently been a number of studies in which truncations of
the new $\omega$-deformed maximal supergravity have been made, typically
with the focus being on finding scalar-field truncations in which the
scalar potential still has a non-trivial dependence on the parameter $\omega$
\cite{Borghese:2012qm,Borghese:2012zs,Borghese:2013dja,Blaback:2013sda,Guarino:2013gsa, Tarrio:2013qga,Anabalon:2013eaa}.
This can lead to a richer structure of anti-de Sitter (AdS) stationary
points and domain-wall solutions,
with the nature of the vacuum state now being dependent on $\omega$.

    It is also of interest to study truncations of the new ${\cal N}=8$
supergravities that give rise to families of supergravities with reduced
supersymmetry.  In this paper, we investigate the consistent embedding of
${\cal N}=2$ gauged $U(1)^4$ STU supergravity, and also the embedding of
${\cal N}=4$ gauged $SO(4)$ supergravity.\footnote{In the
corresponding ungauged theory, the term STU refers to the triality of
$SL(2,R)\times SL(2,R)\times SL(2,R)$ global symmetries, and was
extensively discussed in \cite{dulura}.  In the gauged STU model, the
global symmetry in the bosonic sector
is reduced to $U(1)\times U(1)\times U(1)$.  However, since one of the
factors in the symmetry group is an electric/magnetic duality symmetry,
this factor is broken completely in the full STU gauged supergravity,
because of the minimal coupling of the fermions to the gauge fields.
Further details of the gauged STU supergravity can be found in
\cite{duffliu,tenauthor}.}
In both cases, with the embedding we
consider, we find that the scalar potential of the truncated theory no
longer carries any dependence on the deformation parameter of the larger
${\cal N}=8$ theory, even though the ${\cal N}=8$ potential does, of course,
depend upon $\omega$.  However, this does not mean that the entire bosonic
sector
of the truncated theory is necessarily independent of $\omega$.  In the
case of the truncation to the $U(1)^4$ gauged STU models, we find that the
scalar couplings in the gauge-field kinetic terms carry non-trivial
$\omega$ dependence, and so, for example, charged black hole solutions
will, in general, acquire $\omega$-dependent modifications.

    If one stays purely within the bosonic sector, the $\omega$ dependence 
of the scalar couplings to the gauge-field kinetic terms can be absorbed
by means of duality transformations, together with scalar field redefinitions.
However, since the gauge potentials have minimal couplings to the
fermions, this means that the necessary duality transformations are
not in fact symmetries of the full supergravity theory, and so
$\omega$ is a non-trivial parameter of the full $\omega$-deformed STU
supergravity theories.  In particular, it can affect the supersymmetry
and the fermion couplings.

    We also investigate some
further consistent truncations, where the resulting theories are considerably
simplified.  One obvious such example is where one sets the four $U(1)$
gauge fields pairwise equal.  At the same time, for consistency, four of
the six scalar fields are set to zero, leaving just a single dilaton/axion
pair.  In this truncation it turns out that the $\omega$ dependence in the
scalar couplings of the kinetic terms for the remaining two gauge fields
can in fact be removed merely by using a shift symmetry of the axionic scalar,
with no need to perform any dualisation of the gauge fields,
and so then the entire theory becomes independent of the $\omega$
parameter.

     This pairwise equal truncation can in fact be viewed as
the abelian $U(1)\times U(1)$ truncation of ${\cal N}=4$ gauged $SO(4)$
supergravity.  It is of interest therefore to investigate the
truncation of the deformed ${\cal N}=8$ supergravity to ${\cal N}=4$.  For
the embedding we adopt, we find that the same thing happens; namely, that
the scalar potential is independent of $\omega$ and furthermore that
the $\omega$-dependence in the scalar couplings to the gauge-field kinetic
terms can be eliminated by means of a shift symmetry transformation of the
axionic scalar.  Interestingly, therefore, although one does not obtain
a family of deformed ${\cal N}=4$ gauged $SO(4)$ supergravities via
truncation from the deformed ${\cal N}=8$ supergravities by this method, there
do in fact exist deformed ${\cal N}=4$ gauged $SO(4)$ supergravities, as
constructed by de Roo and Wagemans long ago \cite{deroowag}.  The
deformation parameter $\alpha$ in these theories is very similar in
nature to the deformation parameter $\omega$ in the gauged
${\cal N}=8$ theory, being associated
with a parameterisation of the duality complexion of the gauge fields prior to
gauging.  It may be, however, that the deformations in the ${\cal N}=4$
theories are intrinsic to the ${\cal N}=4$ gaugings, and they do not
necessarily have a larger interpretation with ${\cal N}=8$.

   Returning to the deformed $U(1)^4$ gauged STU models, we find that
different further truncations are possible for which the $\omega$
deformation parameter remains non-trivial.  For example, we can 
perform a ``$1+3$'' truncation in which three of the original gauge fields
are set equal, giving rise to a theory with just two remaining independent
gauge fields with $A_\mu^{(1)}=A_\mu^1$ and $A_\mu^{(2)}=A_\mu^{(3)}= 
A_\mu^{(4)}=A^2_\mu/\sqrt3$.  At the same time, for consistency, 
the three dilatonic scalars of the STU
theory are equated, and the three axionic scalars are equated.
The resulting supergravity theory is in fact a gauged version of
the Poincar\'e supergravity one would obtain by reducing five-dimensional
minimal supergravity on a circle.  It can easily be seen that the $\omega$
parameter enters non-trivially in the scalar couplings to the gauge field
kinetic terms, since the duality transformation that would remove it lies 
outside the symmetry group of the scalar coset.  Thus since this
duality transformation can no longer be performed in the gauged theory,
where the gauge potentials couple minimally to the fermions, we
have a non-trivial family of $\omega$-deformed gauged supergravities in
this truncation.  

   A further consistent truncation allows the second gauge field
$A_\mu^2$ and the axionic scalar
to be set to zero.  Of course, this is no longer a supersymmetric theory, but
it is a consistent truncation of the STU supergravity. 
The resulting bosonic theory is the $\omega$-deformed
generalisation of the ``Kaluza-Klein theory'' obtained by the circle
reduction of pure five-dimensional Einstein gravity, augmented by the addition
of the scalar potential.  The $\omega$ parameter is again non-trivial, in
the sense that it can only be removed by performing a dualisation of
the remaining gauge field, which would be disallowed because its gauge 
potential couples minimally to the fermions, and so viewed as a
truncation of the 
full STU model in which fermions are retained, it 
will still carry the imprint of the $\omega$ parameter.

   The paper is organised as follows.  In section 2, we describe the 
truncation of the fields of ${\cal N}=8$ supergravity to those of
${\cal N}=2$ STU supergravity.  Substituting into the formalism
of the $\omega$-deformed ${\cal N}=8$ theories, we thereby obtain the
corresponding $\omega$-deformed STU supergravity theories.  Our focus
is on the bosonic sector, together with finding the terms in the
supersymmetry transformation rules for the fermionic fields that are
needed in order to investigate the supersymmetry of bosonic solutions.
We also determine the range of the $\omega$ parameter in the STU supergravities
that spans the space of inequivalent models.  In section 3 we examine some
further truncations of the $\omega$-deformed STU supergravity theories.
These include the case where  one sets the four gauge fields pairwise
equal, at the same time setting four of the six scalar fields to zero, which,
as we mentioned above, gives a theory where the $\omega$ parameter becomes
trivial.  We also discuss the non-trivial case of the $1+3$ split described
above, where two gauge fields are
retained.  We present dyonic black hole solutions in the further truncation
to a single gauge potential.
In section 4 we examine some features of the family of deformed 
${\cal N}=4$ gauged $SO(4)$ supergravities that were constructed by
de Roo and Wagemans, which do not arise as truncations of the 
$\omega$-deformed ${\cal N}=8$ theories.  After conclusions in section 5, we
include two appendices.  Appendix A contains the detailed form of the
scalar couplings to the gauge-field kinetic terms in the $\omega$-deformed
STU supergravities, and appendix B contains further detailed expressions
of the various tensors that appear in the supersymmetry transformation
rules.

\section{The Embedding of the STU Model}\label{defstu}

\subsection{The bosonic sector}

   The standard gauged STU supergravity theory can be obtained as a
consistent truncation of standard ${\cal N}=8$ gauged $SO(8)$ supergravity.
In this embedding, the $SO(8)$ gauge fields are truncated
to retain only those in the $U(1)^4$ Cartan subalgebra
of $SO(8)$.  As well as the metric and the four $U(1)$ gauge fields,
the bosonic sector
includes three scalars from the ${\bf 35}_v$ of
scalar fields, and three axionic scalars from the ${\bf 35}_c$ of
pseudoscalar fields in the gauged $SO(8)$ theory.  The fermions will be 
discussed in section 2.3.

The embedding was described in \cite{duffliu} in the case of a further
truncation
in which the three axionic scalar fields are set to zero.\footnote{This
further truncation would be inconsistent in general, in the sense that the
equations of motion for the axionic scalars would not permit setting
them to zero in generic solutions.  However, if one restricts attention
to solutions where the wedge products of pairs of field strengths vanishes,
then the axions can be set to zero.}  The complete
embedding of the bosonic sector of the gauged STU model, including also
the axions, was given in \cite{tenauthor}.  This was achieved by
giving an explicit form for the 56-bein
\be
{\cal V} = \begin{pmatrix} u_{ij}{}^{IJ} & v_{ijKL}\cr
              v^{k\ell IJ} & u^{k\ell}{}_{KL}
            \end{pmatrix}\,,\label{Vuv}
\ee
that enters in the construction of ${\cal N}=8$ supergravity given in
\cite{dewitnic}.  In order to construct the corresponding $\omega$-deformed
version of the gauged STU supergravity model, we may simply take the
same construction of the ${\cal V}$ matrix, and substitute it into
the formulae presented in \cite{dewitnicnew} for the $\omega$-deformed
${\cal N}=8$ supergravity.

  We shall use the same notation and conventions as in \cite{duffliu} and
\cite{tenauthor}.  In \cite{tenauthor},
${\cal V}$ was expressed in the symmetric
gauge
\be
{\cal V} =\exp\Big\{ -\fft1{2\sqrt2} \begin{pmatrix} 0 & \phi_{ijk\ell}\cr
     \phi^{mnpq} & 0
      \end{pmatrix}\Big\}\,,\label{Vdef}
\ee
where $\phi^{ijk\ell}$ is totally antisymmetric and self-dual in the sense that
\be
\bar \phi^{ijk\ell} = \phi_{ijkl}= \fft1{4!} \ep_{ijk\ell mnpq}
\, \phi^{mnpq}\,.
\ee
Denoting the index pairs $\{12, 34, 56, 78\}$ by $\{\alpha\}$ with
$\alpha=1,2,3,4$, the self-dual tensor $\phi^{ijk\ell}$ takes the form
\be
\phi^{ijk\ell}=\phi_{ijk\ell} =
\sqrt2 [(\Phi^{(1)}\ep^{(12)} + \bar\Phi^{(1)}\ep^{(34)})  +
        (\Phi^{(2)}\ep^{(13)} + \bar\Phi^{(2)}\ep^{(24)})  +
        (\Phi^{(3)}\ep^{(14)} + \bar\Phi^{(3)}\ep^{(23)})]\,,
\label{phidef}
\ee
where $\ep^{(12)}_{ijk\ell}=\pm1$ whenever $\{i,j,k,\ell\}$ is an even (odd)
permutation of $\{1,2,3,4\}$, $\ep^{(13)}_{ijk\ell}=\pm1$ whenever
$\{i,j,k,\ell\}$ is an even (odd)
permutation of $\{1,2,5,6\}$, and so on.
We write the three complex scalars as
\be
\Phi^{(a)}= \phi_a\, e^{\im \sigma_a}\,.
\ee

   The scalar kinetic terms are constructed as $-\ft1{48}
{\cal A}_{\mu}^{ijk\ell} {\cal A}^\mu_{ijk\ell}$, where
 ${\cal A}_\mu^{ijk\ell}$ is given by \cite{dewitnic}
\be
D_\mu{\cal V}\, {\cal V}^{-1}= -\fft1{2\sqrt2} \,
\begin{pmatrix} 0 & {\cal A}_\mu^{ijk\ell}\cr
  {\cal A}_{\mu\, ijk\ell} & 0\end{pmatrix}\,.\label{cAdef}
\ee
These kinetic terms are independent of the $\omega$ deformation of
the ${\cal N}=8$ theory.

The scalar potential $V$ is defined in the standard way in ${\cal N}=8$ gauged
$SO(8)$ supergravity, except that now with the $\omega$ deformation the
definition of the $T$ tensor is modified.  Thus one has \cite{dewitnicnew}
\be
V = g^2 \Big[-\ft34 |A_\1{}^{ij}|^2 +
        \ft1{24} |A_{\2\, i}{}^{jk\ell}|^2\Big]\,,
\label{pot}
\ee
where $T_i{}^{jk\ell}$ is decomposed into its irreducible parts
\be
T_i{}^{jk\ell} = -\ft32 A_\1{}^{j[k}\, \delta^{\ell]}{}_i -
    \ft34 A_{\2\, i}{}^{jk\ell}\,.
\ee
Here $A_\1{}^{ij}$ is symmetric in $ij$ and $A_{\2\, i}{}^{jk\ell}$ is
antisymmetric in $jk\ell$ and traceless, i.e. $A_{\2\, i}{}^{ik\ell}=0$.
The $T$ tensor in the $\omega$-deformed theory is defined by
\cite{dewitnicnew}
\be
T_i{}^{jk\ell}= (e^{-\im\omega}\, u^{k\ell}{}_{IJ} +
e^{\im\omega}\, v^{k\ell IJ})(u_{im}{}^{JK}\, u^{jm}{}_{KI} -
  v_{im JK}\, v^{jmKI})\,.
\ee

The four gauge fields of the deformed STU model are taken to be in the
$U(1)^4$ Cartan subalgebra of the $SO(8)$ gauge fields, whose field
strengths are $F^{IJ}= dA^{IJ} - \ft12 g A^{IK}\wedge A^{KJ}$.  Thus,
taking into account an $SO(8)$ triality rotation as discussed in
\cite{duffliu}, we write
\bea
A^{12}&=&\ft1{2}\, [A^{(1)} + A^{(2)} + A^{(3)} + A^{(4)}]  \,,\quad
A^{34}= \ft1{2}\, [A^{(1)} + A^{(2)} -A^{(3)} - A^{(4)}] \,,\nn\\
A^{56}&=&\ft1{2}\, [A^{(1)} - A^{(2)} + A^{(3)} - A^{(4)}]  \,,\quad
A^{78}= \ft1{2}\, [A^{(1)} - A^{(2)} -A^{(3)} + A^{(4)}] \,.
\label{Acombs}
\eea
The corresponding field strengths $F^{(\alpha)}$ will simply be given by
\be
F^{(\alpha)}=dA^{(\alpha)}\,,\qquad \alpha=1,2,3,4\,.
\ee
The kinetic terms for the gauge fields are calculated from those of
the gauge fields in the $\omega$-deformed $SO(8)$ gauged supergravity,
using the expressions given in \cite{dewitnicnew}.  Specifically, one
has $e^{-1}\, {\cal L}_F = 
   -\ft18 (\im G^{+\, \mu\nu}_{IJ} \, F^{+\, IJ}_{\mu\nu} +
\hbox{h.c.})$, where\footnote{Note that in (\ref{Gdef}), and elsewhere also, 
we are not taking into account
higher-order fermion terms.  Our object in this paper is to obtain the bosonic
sector of the $\omega$-deformed STU supergravity theories, and also to 
obtain those terms in the supersymmetry transformation rules that are
sufficient for testing the supersymmetry of bosonic solutions.}
\be
\im\, (u^{ij}{}_{IJ} + e^{2\im\omega}\, v^{ijIJ})G^+_{\mu\nu IJ}=
(u^{ij}{}_{IJ} - e^{2\im\omega}\, v^{ijIJ}) F^{+}_{\mu\nu}{}^{IJ}\,,
\label{Gdef}
\ee
and where $F^+_{\mu\nu}$ is defined for any 2-form as the complex self-dual
projection\footnote{We are using conventions similar to those
in \cite{freevanp} here, in which $G_{\mu\nu} =
  \delta {\cal L}/\delta F_{\mu\nu}$ and hence $G^+_{\mu\nu} = Z\, F^+_{\mu\nu}
= \im M F^+_{\mu\nu}$.  (Our definitions of the self-dual and anti-self-dual
projections, $F^\pm_{\mu\nu}=\ft12 (F_{\mu\nu} \pm 
  \im\, {^*\!F}_{\mu\nu})$, with
${^*\!F}_{\mu\nu}=\ft12 \epsilon_{\mu\nu\rho\sigma} F^{\rho\sigma}$, 
are opposite
to those in \cite{freevanp}.)}

\be
F^+_{\mu\nu}= \ft12( F_{\mu\nu} + \ft12\im\, \epsilon_{\mu\nu\rho\sigma}
   F^{\rho\sigma})\,.
\ee

   Using the expressions for the $u$ and $v$ tensors that
we can read off from (\ref{Vuv}) and (\ref{Vdef}), we are now in a position
to calculate the bosonic Lagrangian for the $U(1)^4$ truncation of the
$\omega$-deformed $SO(8)$ supergravity.  We find that the scalar potential
 (\ref{pot}) is unmodified from its expression in the original $\omega=0$
theory.  The
bosonic Lagrangian for the $\omega$-deformed STU model is then given by
\be
e^{-1}\, {\cal L} = R
  -\ft12  \sum_{a=1}^3 \Big( (\del\phi_a)^2 +
  \sinh^2\phi_a\, (\del\sigma_a)^2\Big) -
\ft14 \Big(F_{\mu\nu}^{+\, (\alpha)}\, {\cal M}_{\alpha\beta} F^{+\, (\beta)
\mu\nu}
+  \hbox{h.c.}\Big) - V\,,\label{defstulag}
\ee
where the scalar potential is
\be
V = -2 g^2(\cosh\phi_1 + \cosh\phi_2 + \cosh\phi_3)\,.
\ee
The complete expression for ${\cal M}_{\alpha\beta}$, which is
rather complicated, is given in appendix A.

   The important point to note is that although the scalar kinetic
terms and potential are unmodified by the $\omega$ deformation, the
kinetic terms for the gauge fields, encapsulated in the matrix
${\cal M}_{\alpha\beta}$, depend upon $\omega$ in a non-trivial way.
Specifically, for no choice of the constants $k_a$ is it possible 
to use the residual $U(1)^3$ global
symmetry $\sigma_a\rightarrow \sigma_a + k_a$  of the scalar kinetic and 
potential terms to absorb the $\omega$ dependence in the 
matrix ${\cal M}_{\alpha\beta}$.   Of course, in the {\it ungauged} theory,
the $\omega$ parameter could be absorbed everywhere if one made an
appropriate electric/magnetic duality transformation on the field 
strengths.  However, this transformation would not be contained
within the $SL(2,R)^3$ global symmetry of the ungauged theory.\footnote{If
it {\it were} contained within the $SL(2,R)^3$ global symmetry, then
it would have been possible to absorb the $\omega$ parameter by means
of a field redefinition of the scalars.} (See appendix A for details.) In the 
purely bosonic sector of the gauged theory, the kinetic terms of the field
strengths are identical to those in the ungauged theory, and hence the
same duality transformation would allow one to remove the 
$\omega$ dependence here as well.  However, in the complete gauged
theory including the fermions, the minimal coupling of the spinors to
the gauge potentials prevents one from making the duality transformation,
and hence $\omega$ cannot be removed by means of field redefinitions in
the $\omega$-deformed STU supergravities.

\subsection{Range of the $\omega$ parameter}

   Having obtained the embedding of the STU model into the $\omega$-deformed
${\cal N}=8$ gauged supergravity, the question arises as to what is the
range of the angle $\omega$ that parameterises inequivalent STU
supergravity theories.  In the full ${\cal N}=8$ theory the inequivalent
gaugings are parameterised by $\omega$ lying in the interval
$0\le \omega\le \ft18\pi$ \cite{dalinvtri,dewitnicnew}.  
The arguments that show that the ``naive'' answer of $0\le\omega<2\pi$
for the parameter range of inequivalent theories is actually reduced to
$0\le \omega\le \ft18\pi$ depend quite subtly upon the use of field
redefinitions involving the ${\cal N}=8$ fields.
In the truncation to
the ${\cal N}=2$ STU theories, many of the original fields are set to
zero, and so it is necessary to investigate within the restricted set
of fields that are retained to see whether or not the conclusions about the
non-trivial range of the $\omega$ parameter remains unchanged.

   Looking at the form of the matrix of scalar fields
${\cal M}_{\alpha\beta}$ that appears in the gauge-field kinetic terms in
(\ref{defstulag}), and which is given in appendix A, it is evident that
there is a symmetry under which we send
\be
\omega \longrightarrow \omega + \fft{\pi}{2}\,,\qquad \phi_a
\longrightarrow -\phi_a\,,\label{omegasym1}
\ee
or in other words, ${\cal M}_{\alpha\beta}(\phi_a,\sigma_a,\omega)=
{\cal M}_{\alpha\beta}(-\phi_a,\sigma_a,\omega+\ft12\pi)$.  This
corresponds to the first of the three discrete symmetries discussed in 
\cite{dewitnicnew}, and as can be seen from (\ref{uexp}) and (\ref{vexp}),
we have $u_{ij}{}^{KL}\longrightarrow u_{ij}{}^{KL}$ and 
$v_{ijKL}\longrightarrow - v_{ijKL}$.  
The matrix ${\cal M}_{\alpha\beta}$, which is symmetric, also has
the property
\be
{\cal M}^*_{\alpha\beta}(\phi_a,\sigma_a,\omega)= {
\cal M}_{\alpha\beta}(\phi_a,-\sigma_a,-\omega)\,,
\ee
where the star denotes a complex conjugation.  The complex conjugation
of ${\cal M}_{\alpha\beta}$ has the effect in the Lagrangian (\ref{defstulag})
of reversing the sign of the
$\epsilon^{\mu\nu\rho\sigma}\, F^{(\alpha)}_{\mu\nu}
  F^{(\beta)}_{\rho\sigma}$ terms, while
leaving the sign of the actual kinetic terms $F^{(\alpha)}_{\mu\nu}\,
   F^{(\beta)\, \mu\nu}$ unchanged.  This sign change can be undone by
means of a parity reversal, and this is the third of the three discrete 
symmetries discussed in \cite{dewitnicnew}.  Finally, we consider the
second discrete symmetry considered in \cite{dewitnicnew}, namely 
\be
\omega\longrightarrow \omega + \fft{\pi}{4}\,.\label{pi/4}
\ee
This involved a discrete $SU(8)$ matrix $e^{\im \pi/8}\, P_8$, where $P_8$ is
a real and orthogonal $8\times 8$ matrix with $\det(P_8)=-1$, and satisfying
$P_8^2=1$.  An example that is convenient for our purposes is
\be
P_8 =\hbox{diag}\, (-1,1,1,1,1,1,1,1)\,.
\ee
 From the transformations in equation (2.2) of \cite{dewitnicnew}, this will
give a symmetry of the theory if $u_{ij}{}^{KL}$ reverses sign 
when exactly one of the four indices is a ``1,'' and if $v_{ijKL}\longrightarrow
e^{\im \pi/2}\, v_{ijKL}$ for all index assignments where exactly one
index is a ``1,'' and $v_{ijKL}\longrightarrow
e^{-\im \pi/2}\, v_{ijKL}$ otherwise.  It can be seen from (\ref{uexp}) and
(\ref{vexp}) that these  can be implemented by means of the transformations
\be
\sigma_a \longrightarrow \sigma_a'= \sigma_a -\fft{\pi}{2} \label{sigmatrans}
\ee
of the three axionic scalars of the truncation to the STU model.
The symmetry also requires sending \cite{dewitnicnew}
\be
A_\mu^{12}\longrightarrow -A_\mu^{12}\,,\quad
(A_\mu^{34}, A_\mu^{56},A_\mu^{78})\longrightarrow 
  (A_\mu^{34}, A_\mu^{56},A_\mu^{78})\,.\label{Atrans}
\ee
 
 The upshot of these symmetries 
is that the $\omega$-deformed STU theories may be 
viewed as inequivalent if 
$\omega$ lies in the interval
\be
0\le \omega \le \fft{\pi}{8}\,,\label{omegarange}
\ee
just as in the case of the $\omega$-deformed ${\cal N}=8$ 
supergravities.\footnote{We are very grateful to the referee for suggesting
to us that the deformed STU theories might also be equivalent under the
transformation (\ref{pi/4}).}

\subsection{Supersymmetry of the $\omega$-deformed STU supergravities}

   Here, we shall present the terms in the supersymmetry transformation
rules for the $\omega$-deformed STU supergravities that are relevant for
investigating the supersymmetry of bosonic solutions in the theories.
We obtain these by substituting the
STU truncation into the transformation rules of the $\omega$-deformed
${\cal N}=8$ theory, which are given in \cite{deWit:2007mt}.  Up to
the order in fermion fields to which we shall be working, they are 
given by 
\begin{eqnarray}
  \label{eq:susy-transformations-gauged}
  \delta \psi_{\mu}{}^{i}
  &=&
  2\,\mathcal{D}_{\mu}\epsilon^{i}
  +\ft{1}{4}\,\hat{\mathcal{H}}^{-}_{\rho\sigma}{}^{ij}\,
  \gamma^{\rho\sigma}\gamma_{\mu}\epsilon_{j}  
  +g \,A_\1{}^{ij}\,\gamma_{\mu}\,\epsilon_{j} +\cdots\,,
     \nonumber\\
  \delta \chi^{ijk}
  &=&
  -\,\hat{\cal A}_{\mu}^{ijkl}\,\gamma^{\mu}\epsilon_{l}
  + \ft3{2\sqrt2} \, \hat{\mathcal{H}}^{-}_{\mu\nu}{}^{[ij}
  \gamma^{\mu\nu}\epsilon^{k]}
  - \sqrt2\, g\,A_{\2 l}{}^{ijk}\,\epsilon^{l} +\cdots\,,   \nonumber\\
  \delta e_{\mu}{}^{a}&=&
  \bar\epsilon^{i}\gamma^{a}\psi_{\mu i} ~+~
  \bar\epsilon_{i}\gamma^{a}\psi_{\mu}{}^i \,, \nonumber\\
    \delta A_{\mu}{}^{IJ}
    &=&
    - \sqrt2 (e^{i\omega}u_{ij}^{~IJ}+e^{-i\omega}v_{ijIJ})\,\Big(
    \bar\epsilon_{k}\,\gamma_{\mu}\,\chi^{ijk}
    +2\sqrt{2}\, \bar\epsilon^{i}\,\psi^j_{\mu}\Big)~+~ {\rm h.c.}
    \,,\nn\\
 \delta u^{ij}_{IJ} &=&  -2\sqrt{2}\,v_{klIJ} \, \Big(
  \bar\epsilon^{[i}\chi^{jkl]}+\ft1{24}\varepsilon^{ijklmnpq}\,
  \bar\epsilon_{m}\chi_{npq}\Big)   \,,  \nonumber \\
 \delta v^{ijIJ} &=&  -2\sqrt{2}\,u_{kl}^{~IJ} \, \Big(
  \bar\epsilon^{[i}\chi^{jkl]}+\ft1{24}\varepsilon^{ijklmnpq}\,
  \bar\epsilon_{m}\chi_{npq}\Big)   \,.
\end{eqnarray}
The ellipses in the transformation rules for the fermions represent
terms of higher order in fermion fields.  
The derivation of the complete set of transformation
rules in the ${\cal N}=2$ truncation, 
including all the higher-order fermion terms, would be straightforward
but rather involved.  One also has to take into account the 
compensating transformations that would be necessary
in order to maintain the symmetric gauge choice (\ref{Vdef}) for the 
parameterisation of the coset for the scalar fields \cite{cremjuli}.

In the above equations,
$\mathcal{D}_{\mu}\epsilon^{i}$ is defined to be
\be
\mathcal{D}_{\mu}\epsilon^{i}=
\nabla_{\mu}\epsilon^i+\ft12 {\cal B}^i_{\mu j}\epsilon^j\,,
\label{Bcomp}
\ee
where ${\cal B}^i_{\mu j}$ is the composite and gauge connection,
and $\hat{\mathcal{H}}_{\mu\nu}{}^{ij}$ is defined through
\be
F_{\mu\nu}^{~IJ}=(e^{\im\omega}u_{ij}^{~IJ}+
  e^{-\im\omega}v_{ijIJ})\hat{\mathcal{H}}_{\mu\nu}{}^{ij}\,.\label{Hdef}
\ee
The various tensors $u_{ij}{}^{KL}$, $v_{ijKL}$, $A_{\1}^{ij}$,
$A_{\2\, i}{}^{jk\ell}$, ${\mathcal H}_{\mu\nu}{}^{ij}$, etc., in the
$\omega$-deformed STU model can be found in appendix B.

   In the truncation to the ${\cal N}=2$ STU models, six of the eight
gravitini and eight supersymmetry parameters are set to zero.  Given
our truncation in the bosonic sector, where we keep the four gauge
$U(1)$ potentials $(A_\mu^{12}, A_\mu^{34}, A_\mu^{56}, A_\mu^{78})$,
there are four different ways we could truncate the gravitini and
supersymmetry parameters.  For definiteness, we shall choose the truncation
where we set all $\psi_\mu^i$ and $\epsilon^i$ to zero except
\be
\psi_\mu^i \quad \hbox{and}\quad \epsilon^i\,,\qquad \hbox{for}\ i=1,2\,.
\ee
The corresponding truncation for the spin-$\ft12$ fermions involves setting
all $\chi^{ijk}=0$ except for the six fields
\be
(\chi^{i 34}\,,\chi^{i56}\,, \chi^{i78})\,,\qquad \hbox{for}\ i=1,2.
\ee
One can verify from the complete supersymmetry transformation rules
that the truncation to the ${\cal N}=2$ STU theories is indeed consistent,
in the sense that the variations of the fields that are set to zero
remain equal to zero.
It can be seen from the expressions in (\ref{Bcon}) for the components
of the connection ${\cal B}^i_{\mu j}$ that appears in the 
covariant derivative ${\mathcal D}_\mu$ in (\ref{Bcomp}) that with our
choice for the truncation to ${\cal N}=2$, it is the
potential $A_\mu^{12}$ that is the graviphoton in the supergravity multiplet.
We present the transformation rules for the scalar fields, after taking 
account of the compensating transformations mentioned above, and the
transformation rules for the surviving gauge potentials, at the 
end of appendix B.

\section{Consistent Truncations of the $\omega$-Deformed STU Model}

\subsection{Pairwise equal gauge fields}\label{pairwisesec}

   There is a consistent truncation in which we set the gauge fields pairwise
equal, at the same time truncating out two dilaton/axion pairs.  For
example, we can set
\be
F^{(2)}=F^{(1)}=\fft1{\sqrt2} F\,,\qquad F^{(4)}=F^{(3)}=\fft1{\sqrt2}
\widetilde F\,,\qquad
\phi_2=\phi_3=\sigma_2=\sigma_3=0\,.
\ee
This then leads to the Lagrangian
\bea
e^{-1} {\cal L} &=& R -\ft12 (\del\phi_1)^2 -
  \ft12 \sinh^2\phi_1\, (\del\sigma_1)^2 +2 g^2(2+\cosh\phi_1)\nn\\
&&-
\ft12 \Big[ \fft{1 + z\, e^{2\im\omega}}{1 - z\, e^{2\im\omega}}\,(F^+)^2 +
\fft{1 - z\, e^{2\im\omega}}{1 + z\, e^{2\im\omega}}\, (\widetilde F^+)^2
+\hbox{h.c.}\Big]  \,,
\eea
where
\be
z = e^{\im\sigma_1}\, \tanh\ft12\phi_1\,.
\ee
The electrically charged rotating black hole solutions in this theory 
were obtained in \cite{Chong:2004na}. 
(The general
charged rotating black holes in the ungauged STU supergravity were constructed
in \cite{cvetyoum}.) 
It is evident that in this 
further truncation of the $\omega$-deformed
STU supergravities, the parameter $\omega$ has now become trivial, in the
sense that it can be absorbed into a redefinition of the $\sigma_1$
scalar field,
\be
\sigma_1\longrightarrow \sigma_1 - 2\omega\,.
\ee
We shall discuss the embedding of the nonabelian extension of this theory
to ${\cal N}=4$ gauged $SO(4)$ supergravity, and the related topic of
its deformation that was discovered long ago by de Roo and Wagemans
\cite{deroowag}, in section \ref{deroosec}.

\subsection{$1+3$ split of the gauge fields}\label{13sec}

     In view of the fact that, as we just saw,
the $\omega$ parameter becomes trivial
within the context of the pairwise-equal truncation of the deformed STU
model, it is of interest to see whether there exist different
consistent truncations in which $\omega$ remains non-trivial.
An example is provided by making instead the following truncation.
\bea
\phi_2&=&\phi_3=\phi_1\,,\qquad \sigma_2=\sigma_3=\sigma_1\,,\cr
A_\mu^{(1)} &=& A_\mu^1\,,\qquad A_\mu^{(2)}=A_\mu^{(3)}=A_\mu^{(4)}= 
\fft1{\sqrt3}\, A_\mu^2\,.\label{3plus1trunc}
\eea
If at the same time, in the fermionic sector one sets $\chi^{i34}=\chi^{i56}
=\chi^{i78}$ (with $i=1,2$ as before), the resulting truncated theory
is ${\cal N}=2$ supersymmetric. It is in fact a gauged version of the
Poincar\'e supergravity one obtains by dimensionally reducing five-dimensional
ungauged minimal supergravity on a circle.  The bosonic Lagrangian takes
the form
\be
e^{-1}\, {\cal L}= R - \ft32 (\del\phi_1)^2 -
   \ft32 \sinh^2\phi_1\, (\del\sigma_1)^2  -\ft14 \big( F^{+ a}_{\mu\nu}\,
S_{ab}\, F^{+ a\, \mu\nu} + \hbox {h.c.}\big) -V\,,\label{lag13}
\ee
where 
\be 
V = - 6g^2 \cosh\phi_1
\ee
and the $2\times 2$ matrix $S_{ab}$ has components given by
\bea
S_{11} &=& \fft1{2\Delta}\, 
 \Big[4\cos(2\omega+\sigma_1)\cosh2\phi_1 -(3+ \cos 2\sigma_1)\sinh 2\phi_1 
\cr
&& \qquad\ \  -4\im \sin(2\omega + \sigma_1) \cosh\phi_1
  +    2\im \sin 2\sigma_1 \sinh\phi_1 \Big]\,,\cr
S_{12}&=&S_{21}= - \fft{2\sqrt 3}{\Delta}\, 
  (\im\cos\sigma_1+ \cosh\phi_1\, \sin\sigma_1)\, \sinh\phi_1\sin\sigma_1\,,\cr
S_{22} &=& \fft1{2\Delta}\Big[4\cos(2\omega+\sigma_1)\cosh2\phi_1 +
          (3+ \cos 2\sigma_1)\sinh 2\phi_1\cr
&& \qquad\ \  -4\im \sin(2\omega + \sigma_1) \cosh\phi_1
  -    2\im \sin 2\sigma_1 \sinh\phi_1 \Big]\,,\label{Smatrix}
 \eea
with
\bea
\Delta&=& 2 \cos(2\omega+\sigma_1)\cosh\phi_1 +  2\cos2\sigma_1\sinh\phi_1
-2\im \sin(2\omega+\sigma_1)\cosh2\phi_1\cr
  &&- \im \sin2\sigma_1 \sinh 2\phi_1 \,.\label{Delta}
\eea

   It was observed in \cite{cjlp} that the ungauged theory (and
hence also the scalar and gauge-field kinetic terms in (\ref{lag13})) has
a global $SL(2,R)$ symmetry under which the field strengths $F^1$ and $F^2$
and their duals transform in a 4-dimensional irreducible representation
(see \cite{lumepo} for details). It is evident from the form of the
matrix $S_{ab}$ that the electric/magnetic duality rotation described by
the $U(1)$ subgroup  of this $SL(2,R)$, under which $\sigma_1$ would
shift by the angle of the duality rotation, cannot be used in order
to absorb the parameter $\omega$.  In other words, the duality rotations that
one would have to perform in order to eliminate the parameter $\omega$
from the gauge field kinetic terms in (\ref{lag13}) lie outside the 
global symmetry group of the theory.  

   In fact, it is not difficult to 
work out the explicit form of the duality transformation that removes 
the $\omega$ dependence in the gauge-field kinetic terms.  The general
$Sp(4,R)$ transformations that are symmetries of the gauge field equations 
of motion take the form \cite{freevanp}
\be
\begin{pmatrix} F^1\cr F^2 \cr G_1 \cr G_2\end{pmatrix}
\longrightarrow \Lambda\, 
 \begin{pmatrix} F^1\cr F^2 \cr G_1 \cr G_2\end{pmatrix}\,,\qquad
\Lambda=\begin{pmatrix} A & B\cr C & D\end{pmatrix}\,,\label{Lam}
\ee
where $A$, $B$, $C$ and $D$ are $2\times 2$ matrices satisfying
\be
A^T\, C - C^T\, A=0\,,\qquad B^T\, D - D^T\, B=0\,,\qquad
  A^T\, D - C^T\, B =\oneone\,.
\ee
The required duality transformation will be such that the scalar matrix
$S_{ab}$, viewed as a function of $\omega$, satisfies
\be
\big(C + \im D\, S(\omega)\big)\big(A + \im B\, S(\omega)\big)^{-1} =
        \im S(0)\,.\label{S0S}
\ee
Solving for the matrices $A$, $B$, $C$ and $D$, we find 
$A=D=\oneone \,\cos\omega$ and $B=-C=\oneone\, \sin\omega$, and hence the
transformation matrix $\Lambda$ in (\ref{Lam}) is given by
\be
\Lambda = \begin{pmatrix} \oneone\, \cos\omega &\quad\oneone\, \sin\omega \cr
              -\oneone\, \sin\omega& \quad\oneone\, \cos\omega\end{pmatrix}\,.
\ee
In other words, the two field strengths $F^1$ and $F^2$ are both dualised
in the same way,
\be
F^a \longrightarrow F^a\, \cos\omega + G^a\, \sin\omega\,.
\ee
This $U(1)$ transformation manifestly lies outside the $SL(2,R)$ symmetry 
group of the scalar coset manifold, under which the two gauge fields and
their duals transform as a four-dimensional irreducible representation.
Note that we can write $S(\omega)$ in terms of the undeformed matrix $S(0)$
as
\be
\im\, S(\omega)= (\cos\omega -\im\, S(0)\, \sin\omega)^{-1}\,
(\sin\omega + \im\, S(0)\, \cos\omega)\,.
\ee

  The upshot is that in the gauged
theory where the fermions couple minimally to the gauge potential of the
graviphoton, and hence one therefore cannot perform duality transformations,
the parameter $\omega$ must be non-trivial. 

  It is straightforward to see that $S_{ab}$ given in (\ref{Smatrix}) has
the symmetries
\be
S_{ab}(\phi_1,\sigma_1,\omega)= S_{ab}(-\phi_1,\sigma_1,\omega+\ft12\pi)\,,
\ee
and
\be
S_{ab}^*(\phi_1,\sigma_1,\omega)= S_{ab}(\phi_1,-\sigma_1,-\omega)\,.
\ee
One can also see that the truncation (\ref{3plus1trunc}) is compatible with
the transformations (\ref{sigmatrans}) and (\ref{Atrans}), and so we 
conclude that the range of $\omega$ parameterising inequivalent theories
in this truncation is again given by (\ref{omegarange}).  
In fact one can straightforwardly verify directly 
that the Lagrangian (\ref{lag13}) is invariant under the transformation
$\sigma_1\rightarrow \sigma_1 -\ft12\pi$, combined with a shift of 
$\omega$ by $\ft14\pi$, together with the transformation of the gauge
fields that is implied by (\ref{Atrans}).

\subsection{Single gauge field truncation}\label{1Fsec}

    We may perform a further consistent truncation of the 
$\omega$-deformed supergravity described in section \ref{13sec}, 
theory in which the gauge field $F^2$ and the axion $\sigma_1$ are set 
to zero.   The function $M=S_{11}$ that forms the prefactor of
the remaining gauge-field kinetic term now becomes
\be
M= \fft{\cos\omega  - \im\,e^{3\phi_1}\,\sin\omega}{
   -\im\, \sin\omega +  e^{3\phi_1}\, \cos\omega}\,.\label{Mfun2}
\ee
Within the bosonic theory it is possible to absorb the parameter
$\omega$, by making a $U(1)\in Sp(2,R)$ duality transformation of the field
strength and its dual, 
\be
\begin{pmatrix} F\cr G\end{pmatrix} \longrightarrow
 \Lambda\, \begin{pmatrix} F\cr G\end{pmatrix}\,,\qquad
\Lambda=
            \begin{pmatrix} \cos\alpha &\quad \sin\alpha\cr
                   -\sin\alpha & \quad\cos\alpha\end{pmatrix}\,.
\ee
If we implement this $U(1)$ transformation with the parameter $\alpha=
\omega +\ft12 \pi$,
then from $\im\, M\longrightarrow (c+\im\, d M)(a + \im\, b M)^{-1}$ 
it would
have the effect of replacing $M$ in (\ref{Mfun2}) by
\be
M' = e^{3\phi_1}\,.
\ee
Thus, were it not for the fact that in the full supergravity theory the bare
potential $A_\mu$ appears in the covariant derivatives of the fermions, one
could implement this $U(1)$ duality transformation on this particular
single gauge field truncation of the deformed STU theory where in addition
$\sigma_1$ is set to zero, and thereby
remove the $\omega$ deformation parameter.  This is not possible within the
framework of the supergravity theory, and so in this sense $\omega$ remains
a non-trivial parameter here, even though it is trivial as far as the
bosonic solutions themselves are concerned.  This is very different from the
situation discussed in section \ref{pairwisesec} for the case of
the truncation with pairwise-equal field strengths.  In that case the
$\omega$-parameter was removed purely by means of a constant shift redefinition
of the axionic scalar field.

  The Lagrangian for this $\sigma_1=0$ further truncation of the single gauge
field truncation can be written as
\bea
e^{-1} {\cal L} &=& R -\ft12 (\del\phi)^2 -
 \fft1{4(e^{-\sqrt3\, \phi}\, \sin^2\omega +
         e^{\sqrt3\, \phi}\,\cos^2\omega)}\, F^2 \nn\\
&&-
\fft{\sin2\omega\, \sinh\sqrt3\phi}{8(e^{-\sqrt3\, \phi}\, \sin^2\omega +
         e^{\sqrt3\, \phi}\,\cos^2\omega)}\, \epsilon^{\mu\nu\rho\sigma}\,
   F_{\mu\nu} F_{\rho\sigma} + 6 g^2 \cosh(\fft{\phi}{\sqrt3})\,,
\label{omegatheory}
\eea
where we have now given the remaining scalar field a canonical
normalisation, by defining $\phi=\sqrt3\, \phi_1$.
Note that in
the undeformed case, where $\omega=0$, this becomes the standard Lagrangian
of the gauged ``Kaluza-Klein theory,''\footnote{David Chow has pointed out to
us that the Lagrangian (\ref{omegatheory}) in the special case that
$\omega$ is set equal to 
$\ft14\pi$ was encountered in \cite{dogasowi} as a consistent truncation in
an Einstein-Sasaki reduction of $D=11$ supergravity.}
\be
e^{-1} {\cal L} = R -\ft12 (\del\phi)^2 -\ft14 e^{-\sqrt3\, \phi} \, F^2
+  6 g^2 \cosh(\fft{\phi}{\sqrt3})\,.\label{zerotheory}
\ee
(This is the theory that comes from the Kaluza-Klein reduction of
five-dimensional Einstein gravity, with the added scalar potential.)

   If we stay within the framework of the gauged theory coupled to fermionic
fields, so that one is not allowed to make duality transformations on
the field strength $F_{\mu\nu}$, then it is easy to see that
the bosonic theories described
by (\ref{omegatheory}) are inequivalent for values of the $\omega$
parameter lying in the interval $0\le \omega\le \ft14\pi$. 
This follows from the fact that (\ref{omegatheory}) has a symmetry 
under sending
\be
\omega\longrightarrow \omega + \fft{\pi}{2}\,,\qquad
\phi\longrightarrow -\phi\,.
\ee
Furthermore, if we send $\omega\longrightarrow -\omega$ the sign of the
$\epsilon^{\mu\nu\rho\sigma}\, F_{\mu\nu} F_{\rho\sigma}$ term is reversed,
and this can be undone by means of a parity transformation.  Note that 
because the axionic scalar $\sigma_1$ has been set to zero in this
truncation, we no longer have the further symmetry under $\omega\rightarrow 
\omega + \ft14\pi$ that was present in the STU model, and thus here the
parameter range for inequivalent theories is 
$0\le\omega\le \ft14\pi$ rather than $0\le \omega\le \ft18\pi$.

   As mentioned above, in this single gauge field truncation with
$\sigma_1$ also truncated out, the bosonic solutions are all equivalent
to solutions in the $\omega=0$ theory, if we ignore the minimal
coupling of $A_\mu$ to the fermions and then allow duality
transformations of the field $F_{\mu\nu}$.  Thus
we can construct solutions to the theory (\ref{omegatheory}) by making such
duality rotations on known solutions of (\ref{zerotheory}).  As an example,
we may consider the static dyonic black hole solution of the theory
(\ref{zerotheory}), which was recently constructed in \cite{lupapodyon}.
Implementing the duality rotation we arrive at the
conclusion that the following dyonic black hole solves the equations of motion
coming from (\ref{omegatheory}):
\begin{eqnarray}
ds^2 &=& -(H_1 H_2)^{-\fft12} f dt^2 + (H_1 H_2)^{\fft12} \Big(\fft{dr^2}{f} + r
^2 (d\theta^2 + \sin^2\theta\, d\varphi^2)\Big)\,,\cr
\phi&=&\ft{\sqrt{3}}{2} \log\fft{H_2}{H_1}\,,\qquad
f=f_0 + g^2 r^2 H_1 H_2\,,\qquad f_0=1 - \fft{2\mu}{r}\,,\cr
A&=&\sqrt2\Big[\Big(\fft{(1 - \beta_1 f_0)}{\sqrt{\beta_1\gamma_2}\, H_1}\Big)\,
  \cos\omega -
   \Big(\fft{(1 - \beta_2 f_0)}{\sqrt{\beta_2\gamma_1}\, H_2}\Big)\,
  \sin\omega\Big] \, dt \nn\\
&&+ 2\sqrt 2\mu\,
 \Big[ \fft{\sqrt{\beta_2\gamma_1}}{\gamma_2}\, \cos\omega +
       \fft{\sqrt{\beta_1\gamma_2}}{\gamma_1}\, \sin\omega\Big]\, \cos\theta
d\varphi\,,\cr
H_1&=&\gamma_1^{-1} (1-2\beta_1 f_0 + \beta_1\beta_2 f_0^2)\,,\qquad
H_2=\gamma_2^{-1}(1 - 2\beta_2 f_0 + \beta_1\beta_2 f_0^2)\,,\cr
\gamma_1&=& 1- 2\beta_1 + \beta_1\beta_2\,,\qquad
\gamma_2 = 1-2\beta_2 + \beta_1\beta_2\,.
\label{adsdyon}
\end{eqnarray}
The physical electric and magnetic charges of this black hole are given by
\bea
Q &=& \fft{\mu}{\sqrt2}
\Big(\fft{\sqrt{\beta_1\gamma_2}}{\gamma_1}\, \cos\omega -
\fft{\sqrt{\beta_2\gamma_1}}{\gamma_2}\,  \sin\omega\Big)\,,\nn\\
P &=& \fft{\mu}{\sqrt2}
\Big(\fft{\sqrt{\beta_2\gamma_1}}{\gamma_2}\, \cos\omega +
\fft{\sqrt{\beta_1\gamma_2}}{\gamma_1}\,  \sin\omega\Big)
\eea
and its mass is
\begin{equation}
M= \fft{(1-\beta_1)(1-\beta_2)(1-\beta_1\beta_2) \mu}{\gamma_1 \gamma_2}\,.
\end{equation}
In a similar fashion the purely electric and purely magnetic
rotating black holes in the ``gauged Kaluza-Klein''
theory, which were constructed in \cite{chow1,wu}, can be duality rotated into
solutions of the $\omega$-deformed theory described by (\ref{omegatheory}).
Rotating black holes with charges carried by two of the four gauge fields
of STU supergravity were constructed in \cite{chow2}.  These can
similarly be duality rotated to give solutions in the full $\omega$-deformed 
STU supergravities.

   It is instructive also to examine how the supersymmetry of specific 
solutions depends on the value of the $\omega$ deformation parameter.
For example, within the class of black-hole solutions (\ref{adsdyon}) we
may consider the case where we set $\beta_2=0$ and where we also take
the extremal limit, which is achieved by setting
\be
\beta_1= \fft12  -\fft{\mu}{q}\,,
\ee
and then sending $\mu$ to zero.  This gives
\be
M= \fft14 q\,,\qquad Q=\ft14 q\, \cos\omega\,,\qquad P=\ft14 q\, \sin\omega\,.
\ee
At $\omega=0$, this solution is just a
standard supersymmetric black hole in the gauged STU
supergravity theory, with a single field strength carrying an electric 
charge.  When the deformation parameter is non-vanishing, the solution
is still extremal, with $M=\sqrt{Q^2+P^2}$, but, as one can verify from
the transformation rules (\ref{eq:susy-transformations-gauged}), it
is no longer supersymmetric.  This can most easily be seen by looking
at the gaugino transformation rule, which can be written in the general form
$\delta\chi\sim \Xi\, \epsilon$.  One can verify that the 
determinant of the matrix $\Xi$ has a factor $(1-e^{2\im\omega})$, which is
non-vanishing unless $\omega=0$.  This example suffices to illustrate the
point that although a family of solutions in the $\omega$-deformed STU
supergravities may be obtained from an $\omega=0$ solution merely by means of
a duality transformation, the supersymmetry of the solution is 
nevertheless dependent on the value of the $\omega$ parameter.

Of course, while it can be relatively 
easy to check the supersymmetry of a family of $\omega$-deformed solutions
that are obtained, as above, by means of duality transformations of
previously-known solutions of the undeformed theory, it is likely to
be more challenging to discover new supersymmetric solutions in the deformed
theory that are not related to previously known such solutions in the 
$\omega=0$ theory.

\section{${\cal N}=4$ Gauged $SO(4)$ Supergravities}\label{deroosec}

   The pairwise-equal field strength truncation that we discussed
in section \ref{pairwisesec} can also  be
viewed as an abelian $U(1)\times U(1)$ truncation of ${\cal N}=4$ gauged
$SO(4)$ supergravity. The embedding of the ${\cal N}=4$ theory into the
standard ${\cal N}=8$ gauged $SO(8)$ supergravity was discussed
in detail in \cite{cvlupo}, where the explicit form of the $S^7$
reduction from eleven-dimensional supergravity was also presented. One can
easily verify from the expressions for the $u^{ij}{}^{IJ}$ and
$v_{ijIJ}$ tensors given in \cite{cvlupo} that if one substitutes them into
the $\omega$-deformed ${\cal N}=8$ gauged supergravity, then again the
$\omega$ parameter becomes trivial, since it can be absorbed by
means of a shift transformation of the axionic scalar field.  
In other words, one cannot by this means
derive a non-trivial family of ${\cal N}=4$ gauged $SO(4)$ supergravities
as an embedding in the $\omega$-deformed ${\cal N}=8$ theory.  It is
interesting that nonetheless a family of deformed ${\cal N}=4$ gauged
$SO(4)$ supergravities does exist; it was constructed long ago by
de Roo and Wagemans in \cite{deroowag}.

   The bosonic Lagrangian of the de Roo and Wagemans gauged $SO(4)$ theory,
parameterised by the extra angle $\alpha$, takes the form \cite{deroowag}
\supercramp
\bea
e^{-1}{\cal L} &=& R - \fft{2 \del_\mu Z\, \del^\mu \bar Z}{(1-|Z|^2)^2}
- \ft14\Big[ \fft{1+ Z e^{\im(\alpha-\beta)}}{1- Z e^{\im(\alpha-\beta)}}\,
(F_{(1)}^{+i})^2 + \fft{1+ Z e^{-\im(\alpha+\beta)}}{1- Z e^{-\im(\alpha+\beta)}}\,
(F_{(2)}^{+i})^2 + \hbox{h.c.}\Big] - V\,,\label{drlag}
\eea
\uncramp
where the potential is given by
\be
V= -\fft{1}{(1-|Z|^2)}\, \Big[ (g_1^2 + g_2^2) (1+|Z|^2) -
  |e^{\im\alpha} g_1^2 + e^{-\im\alpha} g_2^2|\,(Z+\bar Z)\Big] -
     4 g_1 g_2 \sin\alpha\,,\label{Vdr}
\ee
and the angle $\beta$ is defined in terms of $\alpha$, $g_1$ and $g_2$ by
\be
e^{\im\beta} = \fft{e^{\im\alpha} g_1^2 + e^{-\im\alpha} g_2^2}{
                   |e^{\im\alpha} g_1^2 + e^{-\im\alpha} g_2^2|}\,.
\label{betadef}
\ee
The standard ${\cal N}=4$ gauged $SO(4)$ supergravity corresponds to
taking $\alpha=\ft12\pi$.

    If we now introduce scalar fields $\varphi$ and $\chi$, and define
\be
Z = \fft{\zeta-1}{\zeta+1}\,,\qquad \zeta = -\im \chi + e^{-\varphi}\,,
\ee
then the scalar kinetic terms in (\ref{drlag}) become $-\ft12(\del\varphi)^2 -
\ft12 e^{2\varphi}\, (\del\chi)^2$ and the potential becomes
\be
V= -\ft12 (g_1^2 + g_2^2) (X^2+\widetilde X^2) -
   \ft12 (g_1^4+g_2^4 + 2 g_1^2 g_2^2 \cos2\alpha)^{1/2}\,
(X^2 - \widetilde X^2) - 4 g_1 g_2 \sin\alpha\,,\label{Vdr2}
\ee
where
\be
X^2= e^{\varphi}\,,\qquad
   \widetilde X^2 = e^{-\varphi} + \chi^2 e^{\varphi}\,.
\ee
One can in fact define new gauge coupling constants, in terms of which
the scalar potential becomes independent of the parameter $\alpha$.
We do this be introducing $\tilde g_1$ and $\tilde g_2$, defined by
\be
\tilde g_1^2 = \ft12 (g_1^2 + g_2^2) +
   \ft12 (g_1^4 + g_2^4 + 2 g_1^2 g_2^2 \cos2\alpha)^{1/2}\,,\quad
\tilde g_2^2 = \ft12 (g_1^2 + g_2^2) -
   \ft12 (g_1^4 + g_2^4 + 2 g_1^2 g_2^2 \cos2\alpha)^{1/2}\,.\label{rels}
\ee
The potential (\ref{Vdr2}) then becomes simply
\be
V= -\tilde g_1^2\, X^2 -\tilde g_2^2\, \widetilde X^2 - 4 \tilde g_1
\tilde g_2\,,
\ee
which is the form for the standard $N=4$ gauged supergravity potential.  In
particular, it is now independent of the de Roo-Wagemans parameter $\alpha$.
Of course, there is a price to be paid, namely that the gauge field
kinetic terms will have a more complicated dependence on $\alpha$, since
the expression (\ref{betadef}) defining $\beta$ must now be written
in terms of $\tilde g_1$ and $\tilde g_2$ rather than $g_1$ and $g_2$.  Also,
the gauge coupling parameters $g_1$ and $g_2$ appearing in the definitions
of the Yang-Mills field strengths, $F^i_{(1)}= dA^i_{(1)} + \ft12 g_1
\ep^i{}_{jk} A^j_{(1)}\wedge A^k_{(1)}$ and
$F^i_{(2)}= dA^i_{(2)} + \ft12 g_2
\ep^i{}_{jk} A^j_{(2)}\wedge A^k_{(2)}$, and in the gauge covariant derivatives
acting on the fermion fields, will now have more complicated expressions in
terms of $\tilde g_1$ and $\tilde g_2$.
Note that from (\ref{rels}) we have
\be
g_1^2 + g_2^2 = \tilde g_1^2 + \tilde g_2^2\,,\qquad
g_1 g_2\, \sin\alpha = \tilde g_1 \tilde g_2\,.
\ee

   An alternative way to write the bosonic sector of the de Roo-Wagemans
theory is to absorb the angle $(\alpha-\beta)$ as a phase factor
in a new complex scalar
\be
z = Z e^{\im(\alpha-\beta)}\,,
\ee
in terms of which (\ref{drlag}) becomes
\bea
e^{-1}{\cal L} &=& R - \fft{2 \del_\mu z\, \del^\mu \bar z}{(1-|z|^2)^2}
- \ft14\Big[ \fft{1+ z}{1-z}\,
(F_{(1)}^{+i})^2 + \fft{1+ z e^{-2\im\alpha}}{1- z e^{-2\im\alpha}}\,
(F_{(2)}^{+i})^2 + \hbox{h.c.}\Big] - V\,,\label{drlag2}
\eea
where now the scalar potential $V$ is written as
\be
V= -\fft1{1-|z|^2}\, \Big(g_1^2 \, \big|1+z\big|^2 + g_2^2\,
\big|1+z e^{-2i\alpha}\big|^2 \Big)- 4g_1g_2\sin\alpha\,.\label{Vwestra}
\ee

  Since the scalar potential in the de Roo-Wagemans theory
is independent of $\alpha$ when written in terms of the redefined gauge
couplings $\tilde g_1$ and $\tilde g_2$ as in (\ref{rels}), then if we
restrict attention to gauge fields in the abelian $U(1)\times U(1)$
subgroup of $SO(4)\sim SU(2)\times SU(2)$, we may then remove all
remaining dependence on the parameter $\alpha$ in the bosonic equations of
motion by performing appropriate dualisations of the abelian gauge
fields, say $F^3_{(1)\, \mu\nu}$ and $F^3_{(2)\, \mu\nu}$, such that the
$e^{\im(\pm\alpha -\beta)}$ phases are removed from the gauge-field
kinetic term prefactors that arise from (\ref{drlag}).\footnote{The
required duality transformations lie outside those contained in the
$SL(2,R)$ global symmetry of the kinetic terms in the Lagrangian, and hence
the phases cannot simply be absorbed by means of scalar field redefinitions.
The situation here is analogous to that for the $\omega$-deformed
STU supergravities, as we discussed in section 2.}
  We can then use the
same kind of technique that we used in section \ref{1Fsec}, to generate
solutions of the deformed family of theories simply by making duality rotations
of already-known solutions of the original undeformed theory.  For example,
a rotating dyonic black hole solution of the pairwise-equal STU gauged
supergravity was recently constructed in \cite{chowcomp}, generalising 
the purely electric rotating black holes \cite{Chong:2004na}.  
It is
then straightforward to construct dyonic solutions in the
de Roo-Wagemans theory, by taking the solution in \cite{chowcomp} with its
field strengths $\bar F_{(1)}$ and $\bar F_{(2)}$, with their duals
$\bar G_{(1)}$ and $\bar G_{(2)}$, and then taking the field strengths in the
de Roo-Wagemans theories to be given by
\bea
F_{(1)} &=& \bar F_{(1)}\, \cos\ft12(\beta-\alpha) -
          \bar G_{(1)}\, \sin\ft12(\beta-\alpha)\,,\nn\\
F_{(2)} &=& \bar F_{(2)}\, \cos\ft12(\beta+\alpha) -
          \bar G_{(2)}\, \sin\ft12(\beta+\alpha)\,,
\eea
with the metric, dilaton and axion fields left unchanged.

\section{Conclusions}

   In this paper we have constructed a family of deformed ${\cal N}=2$
gauged STU supergravities, starting from the recently discovered
family of $\omega$-deformed ${\cal N}=8$ gauged $SO(8)$ supergravities.
The STU theories have a field content comprised of the ${\cal N}=2$
gauged supergravity multiplet coupled to three vector multiplets.  The
four $U(1)$ gauge fields lie in the Cartan subgroup of the original
$SO(8)$ gauge fields of the ${\cal N}=8$ supergravities.

   Unlike some other truncations of the $\omega$-deformed ${\cal N}=8$
supergravities that have been studied recently, in the case of the STU
model truncation the scalar potential is unchanged in the presence of the
$\omega$ parameter.  However, the scalar functions that multiply the
kinetic terms of the $U(1)^4$ gauge fields do depend upon $\omega$ in a
non-trivial way, in the sense that it cannot be absorbed merely by
means of redefinitions of the scalar fields. If one were free to make duality 
transformations of the gauge field strengths also, then the 
$\omega$ parameter could be absorbed. However, the fermions
have minimal couplings to the gauge potentials, and thus in the full
$\omega$-deformed STU supergravity one cannot make duality transformations
in order to absorb the $\omega$ parameter, and so it is in this
sense non-trivial.
Thus although purely bosonic solutions in the $\omega$-deformed
STU supergravities can be rotated into solutions of the usual STU model, 
their supersymmetry properties and their couplings to the fermion
fields are different.

   We then studied two different supersymmetric consistent truncations
of the $\omega$-deformed STU supergravities.  In the first, where
the four gauge fields are set pairwise equal, we arrived at a theory
where the $\omega$ parameter becomes trivial, since it can now be absorbed
by means of a shift symmetry transformation of the axion that remains in
the truncation.  By contrast, in a different consistent truncation in
which again two gauge fields remain, but this time achieved by equating 
three of the original gauge fields, we showed that the $\omega$ deformation
parameter remains non-trivial.  The crucial difference between the two cases
is that in the first, the duality transformation that eliminates the
$\omega$ parameter lies within the global symmetry group of the scalar
coset manifold, and so the same effect can be achieved by performing a
symmetry transformation on the scalar fields, thereby absorbing the 
$\omega$ parameter.  In the second truncation, the required duality 
transformation that would eliminate $\omega$ lies {\it outside} the
global symmetry group of the scalar manifold, and so $\omega$ cannot be
removed by means of scalar symmetry transformations in this case.
Since the duality transformations are disallowed in the gauged
theory because of the minimal couplings of the gauge potentials to
the fermions, the $\omega$ parameter remains non-trivial in this case.

 We presented
a class of static dyonic black holes that generalise some solutions that
were recently constructed in the
usual gauged STU model in \cite{lupapodyon}.
These black holes are embedded within a truncation of the STU theory
in which all except one of the gauge fields are set to zero, at the
same time setting the three dilatonic scalars $\phi_a$ equal, and the
three axionic scalars $\sigma_a$ equal.
Viewed purely as bosonic configurations, the solutions we obtained in
this paper would not be genuinely ``new'' in the sense that we obtained them
by making a duality rotation on the gauge field strength of the previously
obtained solutions. This duality transformation would
be a symmetry relating one member of the family of $\omega$-deformed theories
to another, with different $\omega$,
were it not for the minimal coupling of the gauge potential
to the gravitini.  Thus as solutions of the full deformed STU supergravities,
the dyonic black holes we constructed can be viewed as being new.

   The pairwise-equal truncation of the $\omega$-deformed supergravities,
where, as mentioned above, $\omega$ becomes trivial,  
is in fact itself a $U(1)\times U(1)$
truncation of ${\cal N}=4$ gauged $SO(4)$ supergravity.  This raises the
question as to whether the $\omega$ parameter again becomes trivial if
one embeds the full ${\cal N}=4$ theory into
the $\omega$-deformed ${\cal N}=8$ supergravities.  The embedding for
the ${\cal N}=4$ truncation was given in detail in \cite{cvlupo}, and
using this, we were able to demonstrate that here too, the $\omega$
parameter can be absorbed once the truncation is performed.

    It is intriguing that nevertheless, there does exist a one-parameter
family of deformed ${\cal N}=4$ gauged supergravities with $SO(4)$ gauge
group, constructed in \cite{deroowag}.  We studied some properties
of these theories, and presented some examples of bosonic solutions
that could be obtained by making dualisations of the field strengths
in previously-obtained solutions.

  The discovery of the $\omega$-deformed ${\cal N}=8$ gauged supergravities
has raised many intriguing questions, and opens the way for the investigation
of the solutions and the truncations to smaller theories.  There is
also the important  question
as to whether the deformed supergravities have a higher-dimensional
origin.

\section*{Acknowledgements}

  We are grateful to Andrea Borghese, David Chow,
Hadi Godazgar, Mahdi Godazgar, Henning Samtleben and Oscar Varela
for useful discussions.  The research of H.L. is supported in part by the 
NSFC grants 11175269 and 11235003. The work of Y.P. and C.N.P. is supported 
in part by DOE grant DE-FG02-13ER42020.

\appendix

\section{Gauge Field Terms in $\omega$-Deformed STU Supergravity}

   Here we present the complex matrix ${\cal M}_{\alpha\beta}$ that
gives the kinetic terms for the gauge fields in the $\omega$-deformed
STU supergravity model that we constructed in section \ref{defstu}.  The
matrix can be conveniently written in the form
\be
{\cal M}_{\alpha\beta}= \fft1{D}\, {\cal N}_{\alpha\beta}\,,
\ee
where the denominator $D$ is given by
\be
D = \tilde\alpha_4 e^{-4\im\omega}
   + \tilde\alpha_2 e^{-2\im\omega} + \alpha_0  +
   \alpha_2 e^{2\im\omega} + \alpha_4 e^{4\im\omega}\,,\label{denom}
\ee
with
\bea
\tilde\alpha_4 &=& \ft12(-1+\tc_1^2 + \tc_2^2 + \tc_3^2 +
    2 \tc_1 \tc_2 \tc_3)\,,\qquad
\alpha_4= \ft12(-1+\tc_1^2 + \tc_2^2 + \tc_3^2 - 2 \tc_1 \tc_2 \tc_3)\,,\nn\\
\tilde\alpha_2&=& \ft12 \ts_1 \ts_2 \ts_3
\big[e^{\im(\sigma_1+\sigma_2+\sigma_3)}
  + e^{\im(\sigma_1-\sigma_2-\sigma_3)} +
   e^{\im(-\sigma_1+\sigma_2-\sigma_3)} +
          e^{\im(-\sigma_1-\sigma_2+\sigma_3)}\big]\,,\nn\\
\alpha_2&=& -\ft12 \ts_1 \ts_2 \ts_3\big[e^{-\im(\sigma_1+\sigma_2+\sigma_3)}
  + e^{-\im(\sigma_1-\sigma_2-\sigma_3)} +
   e^{-\im(-\sigma_1+\sigma_2-\sigma_3)} +
          e^{-\im(-\sigma_1-\sigma_2+\sigma_3)}\big]\,,\nn\\
\alpha_0&=& - \ts_1^2 \cos2\sigma_1 - \ts_2^2\cos2\sigma_2-
          \ts_3^2 \cos2\sigma_3\,.
\eea
Here, we have defined
\be
\ts_a=\sinh\phi_a\,,\qquad \tc_a= \cosh\phi_a\,.
\label{double}
\ee
The matrix ${\cal M}_{\alpha\beta}$ is symmetric in its indices
$\alpha$ and $\beta$.

   The diagonal components of the numerator matrix ${\cal N}_{\alpha\beta}$
are given by
\bea
{\cal N}_{11} &=& \tilde \alpha_4 e^{-4\im\omega} - \alpha_4 e^{4\im\omega} +
  (\beta_{01} +\beta_{02} +\beta_{03})\cr
  &&+ (\tilde \beta_{20} +\tilde\beta_{21}+\tilde\beta_{22}
   +\tilde\beta_{23}) e^{-2\im\omega} +
(\beta_{20} +\beta_{21}+\beta_{22} +\beta_{23}) e^{2\im\omega}\,,\cr
{\cal N}_{22} &=& \tilde\alpha_4 e^{-4\im\omega} - \alpha_4 e^{4\im\omega} +
  (\beta_{01} -\beta_{02} -\beta_{03}) \cr
  &&+ (\tilde\beta_{20} +\tilde\beta_{21}-\tilde\beta_{22} -\tilde\beta_{23})
e^{-2\im\omega} +
(\beta_{20} +\beta_{21}-\beta_{22} -\beta_{23}) e^{2\im\omega}\,,\cr
{\cal N}_{33} &=& \tilde\alpha_4 e^{-4\im\omega}- \alpha_4 e^{4\im\omega} +
  (-\beta_{01} +\beta_{02} -\beta_{03}) \cr
  &&+ (\tilde\beta_{20} -\tilde\beta_{21}+\tilde\beta_{22} -\tilde\beta_{23})
e^{-2\im\omega} +
(\beta_{20} -\beta_{21}+\beta_{22} -\beta_{23}) e^{2\im\omega}\,,\cr
{\cal N}_{44} &=& \tilde\alpha_4 e^{-4\im\omega} - \alpha_4 e^{4\im\omega} +
  (-\beta_{01} -\beta_{02} +\beta_{03}) \cr
  &&+ (\tilde\beta_{20} -\tilde\beta_{21}-\tilde\beta_{22} +\tilde\beta_{23})
 e^{-2\im\omega} +
(\beta_{20} -\beta_{21}-\beta_{22} +\beta_{23}) e^{2\im\omega}\,,
\eea
where
\bea
\beta_{01}&=& 2\tc_1 \ts_2 \ts_3 \cos\sigma_2\cos\sigma_3\,,\quad
\beta_{02}= 2\tc_2 \ts_1 \ts_3 \cos\sigma_1\cos\sigma_3\,,\quad
\beta_{03}= 2\tc_3 \ts_1 \ts_2 \cos\sigma_1\cos\sigma_2\,,\nn\\
\tilde \beta_{20}&=&  \tilde\alpha_2\,,\qquad
\beta_{20} = - \alpha_2\,, \\
\tilde\beta_{21}&=& \ts_1 (\tc_2 \tc_3 + \tc_1) \cos\sigma_1\,,\quad
 \tilde\beta_{22}= \ts_2 (\tc_1 \tc_3 + \tc_2) \cos\sigma_2\,,\quad
  \tilde\beta_{23}= \ts_3 (\tc_1 \tc_2 + \tc_3) \cos\sigma_3\,,\nn\\
\beta_{21}&=& \ts_1 (\tc_2 \tc_3 - \tc_1) \cos\sigma_1\,,\quad
 \beta_{22}= \ts_2 (\tc_1 \tc_3 - \tc_2) \cos\sigma_2\,,\quad
  \beta_{23}= \ts_3 (\tc_1 \tc_2 - \tc_3) \cos\sigma_3\,.\nn
\eea

  The off-diagonal components of ${\cal N}_{\alpha\beta}$ are as follows.
\bea
{\cal N}_{12} &=& \Big\{-\ft14 \ts_1 \ts_2 \ts_3
   [e^{\im(\sigma_1+\sigma_2+\sigma_3)}
 +e^{\im(\sigma_1-\sigma_2-\sigma_3)}-
  e^{\im(-\sigma_1+\sigma_2-\sigma_3)} -
   e^{\im(-\sigma_1-\sigma_2+\sigma_3)}]\nn\\
&&+ \im\, \ts_1(\tc_2 \tc_3+\tc_1)\,
         \sin\sigma_1\Big\} e^{-2\im\omega}\nn\\
&&+\Big\{-\ft14 \ts_1 \ts_2 \ts_3 [e^{-\im(\sigma_1+\sigma_2+\sigma_3)}
 +e^{-\im(\sigma_1-\sigma_2-\sigma_3)}-
  e^{-\im(-\sigma_1+\sigma_2-\sigma_3)} -
   e^{-\im(-\sigma_1-\sigma_2+\sigma_3)}]\nn\\
&&+ \im\, \ts_1(\tc_2 \tc_3 -\tc_1)\,
         \sin\sigma_1\Big\} e^{2\im\omega}\nn\\
&&
+\big[\im \,\ts_1^2 \sin2\sigma_1 + 2\tc_1 \ts_2 \ts_3
   \sin\sigma_2\sin\sigma_3]
\eea
and
\bea
{\cal N}_{34} &=& \Big\{-\ft14 \ts_1 \ts_2 \ts_3
[e^{\im(\sigma_1+\sigma_2+\sigma_3)}
 +e^{\im(\sigma_1-\sigma_2-\sigma_3)}-
  e^{\im(-\sigma_1+\sigma_2-\sigma_3)} -
   e^{\im(-\sigma_1-\sigma_2+\sigma_3)}]\nn\\
&&- \im\, \ts_1(\tc_2 \tc_3+\tc_1)\,
         \sin\sigma_1\Big\} e^{-2\im\omega}\nn\\
&&+\Big\{-\ft14 \ts_1 \ts_2 \ts_3 [e^{-\im(\sigma_1+\sigma_2+\sigma_3)}
 +e^{-\im(\sigma_1-\sigma_2-\sigma_3)}-
  e^{-\im(-\sigma_1+\sigma_2-\sigma_3)} -
   e^{-\im(-\sigma_1-\sigma_2+\sigma_3)}]\nn\\
&&- \im\, \ts_1(\tc_2 \tc_3 -\tc_1)\,
         \sin\sigma_1\Big\} e^{2\im\omega}\nn\\
&&
+\big[ \im\, \ts_1^2 \sin2\sigma_1 - 2\tc_1 \ts_2 \ts_3
     \sin\sigma_2\sin\sigma_3]\,,
\eea
with $\{ {\cal N}_{13}, {\cal N}_{24}\}$ being given by making the
cyclic replacements
\be
(\ts_1, \ts_2, \ts_3)\longrightarrow (\ts_2, \ts_3, \ts_1)\,,\quad
(\tc_1, \tc_2, \tc_3)\longrightarrow (\tc_2, \tc_3, \tc_1)\,,\quad
(\sigma_1,\sigma_2,\sigma_3) \longrightarrow
(\sigma_2,\sigma_3,\sigma_1)\label{cycle}
\ee
in $\{ {\cal N}_{12}, {\cal N}_{34}\}$ respectively.  Finally,
$\{ {\cal N}_{14}, {\cal N}_{23}\}$ are obtained by again applying the
cyclic replacements (\ref{cycle}) to  $\{ {\cal N}_{13}, {\cal N}_{24}\}$
respectively.

   It can easily be verified that the deformation parameter $\omega$
cannot be absorbed by making shift transformations of the axial
scalar fields $\sigma_1$, $\sigma_2$ and $\sigma_3$, and hence $\omega$
is a non-trivial parameter in the deformed STU model that we have
constructed.  In fact, it can be seen that the required duality
transformation of the gauge fields that removes the $\omega$ parameter
from the matrix ${\cal M}_{\alpha\beta}$ is
\be
\begin{pmatrix} F^{(\alpha)}\cr G_{(\alpha)}\end{pmatrix}\longrightarrow
 \begin{pmatrix} \oneone\, \cos\omega &\quad \oneone\, \sin\omega\cr
      -\oneone\, \sin\omega &\quad \oneone\, \cos\omega\end{pmatrix}\,
 \begin{pmatrix} F^{(\alpha)}\cr G_{(\alpha)}\end{pmatrix}\,,
\ee
that is to say, a simultaneous duality rotation of all four fields through
the same angle $\omega$.  This element of the $Sp(8,R)$ symmetry of the
gauge-field kinetic terms lies outside the $SL(2,R)^3$ symmetry of the
scalar coset, and hence the $\omega$ parameter is non-trivial here.  It
then follows that we can write the matrix ${\cal M}_{\alpha\beta}$ of
the $\omega$-deformed STU theory in terms of ${\cal M}_{0\, \alpha\beta}$,
the corresponding matrix in the undeformed $(\omega=0$) theory, as
\be
\im\, {\cal M}= (\cos\omega + \im\, {\cal M}_0\, \sin\omega)^{-1}
 \, (\im\, {\cal M}_0\, \cos\omega - \sin\omega)\,.
\ee

\section{Expressions in $\omega$-deformed STU Supergravity}

   Here, we collect some expressions for the various quantities that appear
in the Lagrangian and the supersymmetry transformation rules in the
$\omega$-deformed STU supergravity. The starting point for obtaining these is
the expression for the $56\times 56$ matrix ${\cal V}$, defined by
(\ref{Vdef}) and (\ref{phidef}).  It is useful to note that the
matrices defined by the three terms in (\ref{phidef}), proportional to
$\Phi^{(1)}$, $\Phi^{(2)}$ and $\Phi^{(3)}$, all commute, and thus we
may write ${\cal V}$ is the product of three commuting terms,
\be
{\cal V}= {\cal V}_1\, {\cal V}_2\, {\cal V}_3\,.
\ee
The individual factors are given by evaluating (\ref{Vdef}) with
$\phi_{ijk\ell}$ replaced by
\be
\phi^{(1)}_{ijk\ell}= \sqrt{2}\, (\Phi^{(1)}\, \epsilon^{(12)} +
             \bar\Phi^{(1)}\, \epsilon^{(34)})\,,
\ee
to obtain ${\cal V}_1$, and analogously for ${\cal V}_2$ and ${\cal V}_3$.
From these, we can read off the tensors $u(a)_{ij}{}^{KL}$
and
$v(a)_{ijKL}$, defined from
${\cal V}_a$ for $a=1, 2, 3$ in the obvious way using
(\ref{Vuv}).  This gives\footnote{The $u(a)_{ij}{}^{KL}$
and $v(a)_{ijKL}$ tensors for any specific choice of $a$ can be seen to
coincide with the $u_{ij}{}^{KL}$ and $v_{ijKL}$ given in \cite{cvlupo}
(modulo conventions and correcting a typographical error
in \cite{cvlupo}) for the embedding of ${\cal N}=4$ gauged
$SO(4)$ supergravity into the maximal gauged theory. This should be no
surprise, since the scalar sectors then coincide. What is perhaps less
obvious is that the scalar embedding of the complete STU supergravity
should just be given by a product of the three commuting factors
${\cal V}= {\cal V}_1\, {\cal V}_2\, {\cal V}_3$.}
\bea
u(a)_{\sA\sB}{}^{\sC\sD} &=&  \cosh\ft12\phi_a\, \delta_{\sA\sB}^{\sC\sD}
\,,\quad
u(a)_{\bar\sA\bar\sB}{}^{\bar\sC\bar\sD}
 = \cosh\ft12\phi_a\, \delta_{\bar\sA\bar\sB}^{\bar\sC\bar\sD}\,,\quad
u(a)_{\sA\bar\sB}{}^{\sC\bar\sD}= 
\ft12\delta_\sA^\sC\, \delta_{\bar\sB}^{\bar\sD}
\,,\nn\\
v(a)_{\sA\sB\sC\sD}&=& \ft12\sinh\ft12\phi_a\, e^{-\im\, \sigma_a}\,
 \epsilon_{\sA\sB\sC\sD}\,,\qquad
v(a)_{\bar\sA\bar\sB\bar\sC\bar\sD}= \ft12
 \sinh\ft12\phi_a\, e^{\im\,\sigma_a}\,
 \epsilon_{\bar\sA\bar\sB\bar\sC\bar\sD}\,,
\eea
where we divide the index range $(1,\ldots,8)$ into
$(A=\{1,2,3,4\}\,,\bar A=\{5,6,7,8\})$ when $a=1$;
$(A=\{1,2,5,6\}\,,\bar A=\{3,4,7,8\})$ when $a=2$; and
$(A=\{1,2,7,8\}\,,\bar A=\{3,4,5,6\})$ when $a=3$.

   The complete $u_{ij}{}^{KL}$ and $v_{ijKL}$ tensors for the
matrix ${\cal V}$ can then be seen to be as follows.
The tensor $u_{ij}{}^{KL}$ is given by
\bea
u_{12}{}^{12} &=& u_{34}{}^{34}=u_{56}{}^{56}=u_{78}{}^{78}= \ft12 
                             c_1 c_2 c_3\,,\cr
u_{12}{}^{34}&=& \ft12 c_1 s_2 s_3 /(t_1 t_2)\,,\qquad
u_{34}{}^{12}= \ft12 c_1 s_2 s_3 t_1 t_2\,,\cr
u_{12}{}^{56} &=& \ft12 c_2 s_1 s_3 /(t_1 t_3)\,, \qquad
u_{56}{}^{12} = \ft12 c_2 s_1 s_3 t_1 t_3\,,\cr
u_{12}{}^{78} &=& \ft12 c_3 s_1 s_1 /(t_1 t_2)\,,\qquad
u_{78}{}^{12} = \ft12 c_3 s_1 s_2 t_1 t_2\,,\cr
u_{34}{}^{56} &=& \ft12 c_3 s_1 s_2 t_2/t_1\qquad
u_{56}{}^{34} = \ft12 c_3 s_1 s_2 t_1/t_2\,,\cr
u_{34}{}^{78}&=& \ft12 c_2 s_1 s_3 t_3/t_1\,,\qquad
u_{78}{}^{34} = \ft12 c_2 s_1 s_3 t_1/t_3\,,\cr
u_{56}{}^{78} &=& \ft12 c_1 s_2 s_3 t_3/t_2\,,\quad
u_{78}{}^{56} = \ft12 c_1 s_2 s_3 t_2/t_3\,,\cr
u_{13}{}^{13} &=& u_{14}{}^{14}=u_{23}{}^{23}=u_{24}{}^{24}=
u_{57}{}^{57} = u_{58}{}^{58} = u_{67}{}^{67} = u_{68}{}^{68} =\ft12 c_1\,,\cr
u_{15}{}^{15} &=& u_{16}{}^{16}=u_{25}{}^{25}=u_{26}{}^{26}=
u_{37}{}^{37} = u_{38}{}^{38} = u_{47}{}^{47} = u_{48}{}^{48} =\ft12 c_2\,,\cr
u_{17}{}^{17} &=& u_{18}{}^{18}=u_{27}{}^{27}=u_{28}{}^{28}=
u_{35}{}^{35} = u_{36}{}^{36} = u_{45}{}^{45} = u_{46}{}^{46} =\ft12 c_3\,,
\label{uexp}
\eea
and the tensor $v_{ijKL}$ is given by
\bea
v_{1212}&=& \ft12 s_1 s_2 s_3/(t_1 t_2 t_3)\,,\qquad
v_{3434}= \ft12 s_1 s_2 s_3 t_2 t_3/t_1\,,\cr
v_{5656} &=& \ft12 s_1 s_2 s_3 t_1 t_3/t_2\,,\qquad
v_{7878}= \ft12 s_1 s_2 s_3 t_1 t_2/t_3\,,\cr
v_{1234}&=&v_{3412}= \ft12 c_2 c_3 s_1/t_1\,,\quad
v_{1256}=v_{5612}= \ft12 c_1 c_3 s_2/t_2\,,\quad
v_{1278}=v_{7812}= \ft12 c_1 c_2 s_3/t_3\,,\cr
v_{3456}&=&v_{5634}= \ft12 c_1 c_2 s_3 t_3\,,\quad
v_{3478}=v_{7834}= \ft12 c_1 c_3 s_2 t_2\,,\quad
v_{5678}=v_{7856}= \ft12 c_2 c_3 s_1 t_1\,,\cr
v_{1324}&=& v_{2413}=-v_{1423}=-v_{2314}=-\ft12 s_1/t_1\,,\cr
v_{1526}&=& v_{2615}=-v_{1625}=-v_{2516}=-\ft12 s_2/t_2\,,\cr
v_{1728}&=& v_{2817}=-v_{1827}=-v_{2718}=-\ft12 s_3/t_3\,,\cr
v_{5768}&=& v_{6857}=-v_{5867}=-v_{6758}=- \ft12 s_1 t_1\,,\cr
v_{3748}&=&v_{4837}=-v_{3847}=-v_{4738}=-\ft12 s_2 t_2\,,\cr
v_{3546}&=&v_{4635}=-v_{3645}=-v_{4636}=-\ft12 s_3 t_3\,,\label{vexp}
\eea
where we use the notation
\be
c_a =\cosh\ft12\phi_a\,,\qquad  s_a=\sinh\ft12\phi_a\,,
\qquad t_a= e^{\im\sigma_a}\,.
\ee

  The tensors $A_\1^{ij}$ and $A_{\2\, i}{}^{jk\ell}$ take the forms
\bea
A_\1^{ij}&=&\hbox{diag}\, (a_1,a_1,a_2,a_2,a_3,a_3,a_4,a_4)\,,\\
a_1&=& e^{-\im\omega}\, c_1 c_2 c_3 +
                e^{\im\omega}\, s_1 s_2 s_3 t_1 t_2 t_3\,,\quad
a_2= e^{-\im\omega}\, c_1 c_2 c_3 +
                e^{\im\omega}\, s_1 s_2 s_3 t_1 /(t_2 t_3)\,,\nn\\
a_3&=& e^{-\im\omega}\, c_1 c_2 c_3 +
                e^{\im\omega}\, s_1 s_2 s_3 t_2 /(t_1 t_3)\,,\quad
a_4=e^{-\im\omega}\, c_1 c_2 c_3 +
                e^{\im\omega}\, s_1 s_2 s_3 t_3 /(t_1 t_2)\,,\nn
\eea
and
\bea
A_{\2\, 1}{}^{234} &=& - A_{\2\, 2}{}^{134}=
 -(e^{-\im\omega}\, c_1 s_2 s_3/(t_2 t_3) +
               e^{\im\omega}\, c_2 c_3 s_1 t_1)\,,\cr
A_{\2\, 1}{}^{256} &=& - A_{\2\, 2}{}^{156}=
 -(e^{-\im\omega}\, c_2 s_1 s_3/(t_1 t_3) +
               e^{\im\omega}\, c_1 c_3 s_2 t_2)\,,\cr
A_{\2\, 1}{}^{278} &=& - A_{\2\, 2}{}^{178}=
 -(e^{-\im\omega}\, c_3 s_1 s_2/(t_1 t_2) +
               e^{\im\omega}\, c_1 c_2 s_3 t_3)\,,\cr
A_{\2\, 3}{}^{456} &=& - A_{\2\, 4}{}^{356}=
 -(e^{-\im\omega}\, c_3 s_1 s_2 t_2/t_1 +
               e^{\im\omega}\, c_1 c_2 s_3 /t_3)\,,\cr
A_{\2\, 3}{}^{478} &=& - A_{\2\, 4}{}^{378}=
 -(e^{-\im\omega}\, c_2 s_1 s_3 t_3/t_1 +
               e^{\im\omega}\, c_1 c_3 s_2 /t_2)\,,\cr
A_{\2\, 5}{}^{678} &=& - A_{\2\, 6}{}^{578}=
 -(e^{-\im\omega}\, c_1 s_2 s_3 t_3/t_2 +
               e^{\im\omega}\, c_2 c_3 s_1 /t_1)\,,\cr
A_{\2\, 3}{}^{124}&=& - A_{\2\, 4}{}^{123}=
  -(e^{-\im\omega}\, c_1 s_2 s_3 t_2 t_3 +
     e^{\im\omega}\, c_2 c_3 s_1 t_1)\,,\cr
A_{\2\, 5}{}^{126}&=& - A_{\2\, 6}{}^{125}=
  -(e^{-\im\omega}\, c_2 s_1 s_3 t_1 t_3 +
     e^{\im\omega}\, c_1 c_3 s_2 t_2)\,,\cr
A_{\2\, 5}{}^{346}&=& - A_{\2\, 6}{}^{345}=
  -(e^{-\im\omega}\, c_3 s_1 s_2 t_1/t_2 +
     e^{\im\omega}\, c_1 c_2 s_3 /t_3)\,,\cr
A_{\2\, 7}{}^{348}&=& - A_{\2\, 8}{}^{347}=
  -(e^{-\im\omega}\, c_2 s_1 s_3 t_1 /t_3 +
     e^{\im\omega}\, c_1 c_3 s_2 /t_2)\,,\cr
A_{\2\, 7}{}^{568}&=& - A_{\2\, 8}{}^{567}=
  -(e^{-\im\omega}\, c_1 s_2 s_3 t_2 /t_3 +
     e^{\im\omega}\, c_2 c_3 s_1 /t_1)\,,\cr
A_{\2\, 7}{}^{128}&=& - A_{\2\, 8}{}^{127}=
  -(e^{-\im\omega}\, c_3 s_1 s_2 t_1 t_2 +
     e^{\im\omega}\, c_1 c_2 s_3 t_3)\,.
\eea

  Solving (\ref{Hdef}) gives
the non-zero components ${\bf H}_{\mu\nu} =
(\hat{\mathcal{H}}_{\mu\nu}{}^{12},\hat{\mathcal{H}}_{\mu\nu}{}^{34},
\hat{\mathcal{H}}_{\mu\nu}{}^{56}, \hat{\mathcal{H}}_{\mu\nu}{}^{78})^T$
in terms of the four non-zero components of
$F_{\mu\nu}{}^{ij}$, namely ${\bf F}_{\mu\nu}
=(F_{\mu\nu}{}^{12}, F_{\mu\nu}{}^{34},
F_{\mu\nu}{}^{56}, F_{\mu\nu}{}^{78})^T$, as ${\bf H}_{\mu\nu}={\bf K}\,
{\bf F}_{\mu\nu}$, where $K$ is the $4\times 4$ matrix
\be
{\bf K} = \fft{1}{D^*}\, {\bf J}\,,
\ee
and $D$ is defined in (\ref{denom}), with the star denoting complex
conjugation.  The components of ${\bf J}$ can all be given in terms
of
\bea
{\bf J}_{11} &=& e^{3\im\omega}\,(1+ s_1^2 + s_2^2 + s_3^2)c_1c_2c_3
-e^{\im\omega}\, (c_1^2 t_2 t_3/t_1 + c_2^2 t_1 t_3/t_2 +
   c_3^2 t_1 t_2/t_3)s_1 s_2 s_3 \nn\\
&&-
e^{-\im\omega}\, (s_1^2 t_1^2 + s_2^2 t_2^2 + s_3^2 t_3^2) c_1 c_2 c_3 +
e^{-3\im\omega}\, (c_1^2 +c_2^2+c_3^2 -1)s_1 s_2 s_3 t_1 t_2 t_3\,,\nn\\
{\bf J}_{12}&=& e^{3\im\omega}\, (c_1^2-c_2^2-c_3^2)c_1 s_2 s_3 t_2 t_3 +
  e^{\im\omega}\, (s_2^2 t_2^2 t_1 + s_3^2 t_3^2 t_1 - c_1^2/t_1) s_1 c_2 c_3
\nn\\
&&+ e^{-\im\omega}\, (c_2^2 t_3/t_2 + c_3^2 t_2/t_3 - s_1^2 t_1^2 t_2 t_3)
  c_1 s_2 s_3 +e^{-3\im\omega}\, (s_1^2-s_2^2-s_3^2)s_1 c_2 c_3 t_1\,,
\eea
by defining the cyclic operator ${\cal C}$ and the parity operators
${\cal P}_{12}$, ${\cal P}_{23}$ and ${\cal P}_{13}$ that act on the
scalar fields by
\bea
{\cal C}:&& (\phi_1,\phi_2,\phi_3,\sigma_1,\sigma_2,\sigma_3)\longrightarrow
  (\phi_2,\phi_3,\phi_1,\sigma_2,\sigma_3,\sigma_1)\,,\nn\\
{\cal P}_{12}:&& (\sigma_1,\sigma_2)\longrightarrow (-\sigma_1,-\sigma_2)\,,
\nn\\
{\cal P}_{23}:&& (\sigma_2,\sigma_3)\longrightarrow (-\sigma_2,-\sigma_3)\,,
\nn\\
{\cal P}_{13}:&& (\sigma_1,\sigma_3)\longrightarrow (-\sigma_1,-\sigma_3)\,.
\eea
We then have
\bea
{\bf J}_{22}&=& {\cal P}_{23}({\bf J}_{11})\,,\qquad
{\bf J}_{33}= {\cal P}_{13}({\bf J}_{11})\,,\qquad
{\bf J}_{44}= {\cal P}_{12}({\bf J}_{11})\,,\nn\\
{\bf J}_{13}&=& {\cal C}({\bf J}_{12})\,,\qquad \ \ \
{\bf J}_{14}= {\cal C}({\bf J}_{13})\,,\nn\\
{\bf J}_{21} &=& {\cal P}_{23}({\bf J}_{12})\,,\qquad
{\bf J}_{23} = {\cal P}_{23}({\bf J}_{14})\,,\qquad
{\bf J}_{24} = {\cal P}_{23}({\bf J}_{13})\,,\nn\\
{\bf J}_{31} &=& {\cal P}_{13}({\bf J}_{13})\,,\qquad
{\bf J}_{32} = {\cal P}_{13}({\bf J}_{14})\,,\qquad
{\bf J}_{34} = {\cal P}_{13}({\bf J}_{12})\,,\nn\\
{\bf J}_{41} &=& {\cal P}_{12}({\bf J}_{14})\,,\qquad
{\bf J}_{42} = {\cal P}_{12}({\bf J}_{13})\,,\qquad
{\bf J}_{43} = {\cal P}_{12}({\bf J}_{12})\,.
\label{Jcomps}
\eea

   The scalar field kinetic terms in the Lagrangian (\ref{defstulag})
come from the quantities $\cA_\mu^{ijk\ell}$ that are defined in
(\ref{cAdef}).  We find that they are given by
\bea
\cA_\mu^{1234} &=& \ft1{\sqrt2}(\del_\mu\phi_1 -
       \im \sinh\phi_1\,  \del_\mu\sigma_1)\,
        e^{-\im\sigma_1}\,,\ 
\cA_\mu^{5678} = \ft1{\sqrt2} (\del_\mu\phi_1 +
       \im \sinh\phi_1\, \del_\mu\sigma_1)\,
        e^{\im\sigma_1}\,,\nn\\
\cA_\mu^{1256} &=& \ft1{\sqrt2} (\del_\mu\phi_2 -
       \im \sinh\phi_2\,\del_\mu\sigma_2)\,
        e^{-\im\sigma_2}\,,\ 
\cA_\mu^{3478}= \ft1{\sqrt2} (\del_\mu\phi_2 +
       \im \sinh\phi_2\, \del_\mu\sigma_2)\,
        e^{\im\sigma_2}\,,\nn\\
\cA_\mu^{1278} &=& \ft1{\sqrt2} (\del_\mu\phi_3 -
       \im \sinh\phi_3\, \del_\mu\sigma_3)\,
        e^{-\im\sigma_3}\,,\ 
\cA_\mu^{3456}= \ft1{\sqrt2} (\del_\mu\phi_3 +
       \im \sinh\phi_3\,\del_\mu\sigma_3)\,
        e^{\im\sigma_3}\,.
\eea

   The non-vanishing components of the connection
${\cal B}^i_{\mu j}$ appearing in the covariant derivative
(\ref{Bcomp}) are given by
\bea
{\cal B}^1_{\mu 1}&=&{\cal B}^2_{\mu 2}=
 -\im\, (\partial_{\mu}\sigma_1\sinh^2\ft12\phi_1+
\partial_{\mu}\sigma_2\sinh^2\ft12\phi_2+
\partial_{\mu}\sigma_3\sinh^2\ft12\phi_3 )\,,\nn\\
{\cal B}^3_{\mu 3}&=&{\cal B}^4_{\mu 4}=
-\im\, (\partial_{\mu}\sigma_1\sinh^2\ft12\phi_1-
\partial_{\mu}\sigma_2\sinh^2\ft12\phi_2-
\partial_{\mu}\sigma_3\sinh^2\ft12\phi_3 )\,,\nn\\
{\cal B}^5_{\mu 5}&=&{\cal B}^6_{\mu 6}=
-\im\, (-\partial_{\mu}\sigma_1\sinh^2\ft12\phi_1+
\partial_{\mu}\sigma_2\sinh^2\ft12\phi_2-
\partial_{\mu}\sigma_3\sinh^2\ft12\phi_3 )\,,\nn\\
{\cal B}^7_{\mu 7}&=&{\cal B}^8_{\mu 8}=
-\im\, (-\partial_{\mu}\sigma_1\sinh^2\ft12\phi_1-
\partial_{\mu}\sigma_2\sinh^2\ft12\phi_2+
\partial_{\mu}\sigma_3\sinh^2\ft12\phi_3 )\,,\nn\\
{\cal B}^1_{\mu 2}&=&-{\cal B}^2_{\mu 1}=-gA_\mu^{12}\,,\quad
{\cal B}^3_{\mu 4}=-{\cal B}^4_{\mu 3}=-gA_\mu^{34}\,,\nn\\
{\cal B}^5_{\mu 6}&=&-{\cal B}^6_{\mu 5}=-gA_\mu^{56}\,,\quad
{\cal B}^7_{\mu 8}=-{\cal B}^8_{\mu 7}=-gA_\mu^{78}\,.\label{Bcon}
\eea

   It is useful also to note that although the gauge potentials
that we actually use in the STU model are $A_\mu^{(\alpha)}$, which are
defined in terms of ${\bf A}_\mu\equiv
(A_\mu^{12}, A_\mu^{34}, A_\mu^{56},A_\mu^{78})^T$ by
(\ref{Acombs}), the expression for the gauge-field kinetic terms can be
written more simply in terms of the original fields.  Thus if we also
define ${\bf G}_{\mu\nu}\equiv
 (G_\mu^{12}, G_\mu^{34}, G_\mu^{56},G_\mu^{78})^T$, then the solution
to (\ref{Gdef}) can be written as
\be
{\bf G}_{\mu\nu} = {\bf Q} {\bf F}_{\mu\nu}\,,\qquad
  {\bf Q}= \fft{1}{D}\, {\bf R}\,,
\ee
where the $4\times 4$ matrix ${\bf R}$ has components given by
\bea
{\bf R}_{11} &=&
\ft12 e^{4\im\omega}\, (2 \tc_1\tc_2\tc_3 + 1 -\tc_1^2 -\tc_2^2 -\tc_3^2)
+ \ft12 e^{-4\im\omega}\,(2 \tc_1\tc_2\tc_3 - 1 +\tc_1^2 +\tc_2^2 +\tc_3^2)
\nn\\
&&
+\ft12 e^{2\im\omega}\, [1/(t_1 t_2 t_30 -t_1 t_2/t_3 
      - t_1 t_3/t_2 -t_2 t_3/t_1]
  \ts_1\ts_2\ts_3 \nn\\
&&
+\ft12 e^{-2\im\omega}\, [t_1 t_2 t_3 -t_1 /(t_2 t_3) - t_2 /(t_1 t_3)
                -t_3 /(t_1t_2)]  \ts_1\ts_2\ts_3\nn\\
&&
  +\im\, (\ts_1^2 \sin 2\sigma_1 +
              \ts_2^2 \sin 2\sigma_2 +\ts_3^2 \sin 2\sigma_3 )\,,\\
{\bf R}_{12} &=& e^{2\im\omega}\, (\tc_1-\tc_2-\tc_3)\ts_1/t_1
  - e^{-2\im\omega}\, (\tc_1+\tc_2+\tc_3)\ts_1 t_1 +
   \tc_1\ts_2\ts_3\, \cos(\sigma_2-\sigma_3)\,,\nn
\eea
with the remaining components being given, as with (\ref{Jcomps}), by
\bea
{\bf R}_{22}&=& {\cal P}_{23}({\bf R}_{11})\,,\qquad
{\bf R}_{33}= {\cal P}_{13}({\bf R}_{11})\,,\qquad
{\bf R}_{44}= {\cal P}_{12}({\bf R}_{11})\,,\cr
{\bf R}_{13}&=& {\cal C}({\bf R}_{12})\,,\qquad \ \ \
{\bf R}_{14}= {\cal C}({\bf R}_{13})\,,\cr
{\bf R}_{21} &=& {\cal P}_{23}({\bf R}_{12})\,,\qquad
{\bf R}_{23} = {\cal P}_{23}({\bf R}_{14})\,,\qquad
{\bf R}_{24} = {\cal P}_{23}({\bf R}_{13})\,,\cr
{\bf R}_{31} &=& {\cal P}_{13}({\bf R}_{13})\,,\qquad
{\bf R}_{32} = {\cal P}_{13}({\bf R}_{14})\,,\qquad
{\bf R}_{34} = {\cal P}_{13}({\bf R}_{12})\,,\cr
{\bf R}_{41} &=& {\cal P}_{12}({\bf R}_{14})\,,\qquad
{\bf R}_{42} = {\cal P}_{12}({\bf R}_{13})\,,\qquad
{\bf R}_{43} = {\cal P}_{12}({\bf R}_{12})\,.
\label{Rcomps}
\eea
(Recall that the hatted quantities $\tc_a$ and $\ts_a$ are defined
in (\ref{double}).)
The matrix ${\cal M}_{\alpha\beta}$ that appears in the Lagrangian terms for
the gauge fields in (\ref{defstulag}), and that is presented in appendix A,
is related to ${\bf Q}$ by
\be
{\cal M}_{\alpha\beta}= (\Omega\, {\bf Q}\,\Omega)_{\alpha\beta}\,,
\ee
where
\be
\Omega= \ft12 \begin{pmatrix} 1&1&1&1\cr
                              1&1&-1&-1\cr
                              1&-1&1&-1\cr
                              1&-1&-1&1 \end{pmatrix} = \Omega^{-1}=
             \Omega^T
\ee
is the matrix that implements the change of field variables, $A_\mu^{(\alpha)}=
 (\Omega{\bf A}_\mu)^\alpha$, as in (\ref{Acombs}).

  We also present the supersymmetry transformation rules for the
gauge potentials and the scalar fields of the $\omega$-deformed
STU supergravities, which we discussed in section 2.3.  We find
for the gauge potentials
\bea
\delta A_{\mu}^{12}&=&-\sqrt{2}(e^{\im\omega}c_1s_2s_3t_1t_2
                 +e^{-\im \omega}c_2c_3s_1/t_1)
    \bar\epsilon_{i}\,\gamma_{\mu}\,\chi^{i(1)}\nn\\
    &&-\sqrt{2}(e^{\im \omega}c_2s_1s_3t_1t_3
             +e^{-\im \omega}c_1c_3s_2/t_2)
    \bar\epsilon_{i}\,\gamma_{\mu}\,\chi^{i(2)}\nn\\
    &&-\sqrt{2}(e^{\im \omega}c_3s_1s_2t_1t_2+
             e^{-\im \omega}c_1c_2s_3/t_3)
)\bar\epsilon_{i}\,\gamma_{\mu}\,\chi^{i(3)}\nn\\
    && -2(e^{\im \omega}c_1c_2c_3
                     +e^{-\im \omega}s_1s_2s_3/(t_1t_2t_3))
\bar\epsilon^{i}\,\psi_{\mu}^{ j}\epsilon_{ij}
    ~+~ {\rm h.c.}\,,\nn\\
\delta A_{\mu}^{34}&=&-\sqrt{2}(e^{\im \omega}c_1c_2c_3
               +e^{-\im \omega}s_1s_2s_3t_2t_3/t_1)
    \bar\epsilon_{i}\,\gamma_{\mu}\,\chi^{i(1)}\nn\\
    &&-\sqrt{2}(e^{\im \omega}c_3s_1s_2t_1/t_2
           +e^{-\im \omega}c_1c_2s_3t_3)
    \bar\epsilon_{i}\,\gamma_{\mu}\,\chi^{i(2)}\nn\\
    &&-\sqrt{2}(e^{\im \omega}c_2s_1s_3t_1/t_3
                +e^{-\im \omega}c_1c_3s_2t_2)
)\bar\epsilon_{i}\,\gamma_{\mu}\,\chi^{i(3)}~+~ {\rm h.c.}\,,\nn\\
\delta A_{\mu}^{56}&=&-\sqrt{2}(e^{\im \omega}c_3s_1s_2t_2/t_1
               +e^{-\im \omega}c_1c_2s_3t_3)
    \bar\epsilon_{i}\,\gamma_{\mu}\,\chi^{i(1)}\nn\\
    &&-\sqrt{2}(e^{\im \omega}c_1c_2c_3+
               e^{-\im  \omega}s_1s_2s_3t_1t_3/t_2)
    \bar\epsilon_{i}\,\gamma_{\mu}\,\chi^{i(2)}\nn\\
    &&-\sqrt{2}(e^{\im \omega}c_1s_2s_3t_2/t_3
              +e^{-\im \omega}c_2c_3s_1t_1)
)\bar\epsilon_{i}\,\gamma_{\mu}\,\chi^{i(3)}~+~ {\rm h.c.}\,,\nn\\
\delta A_{\mu}^{78}&=&-\sqrt{2}(e^{\im \omega}c_2s_1s_3t_3/t_1
          +e^{-\im\omega}c_1c_3s_2t_2)
    \bar\epsilon_{i}\,\gamma_{\mu}\,\chi^{i(1)}\nn\\
    &&-\sqrt{2}(e^{\im \omega}c_1s_2s_3t_3/t_2
             +e^{-\im \omega}c_2c_3s_1t_1)
    \bar\epsilon_{i}\,\gamma_{\mu}\,\chi^{i(2)}\nn\\
    &&-\sqrt{2}(e^{\im \omega}c_1c_2c_3
                   +e^{-\im \omega}s_1s_2s_3 t_1t_2/t_3
)\bar\epsilon_{i}\,\gamma_{\mu}\,\chi^{i(3)}~+~ {\rm h.c.}\,,
\eea
where $\chi^{i(1)}$, $\chi^{i(2)}$ and $\chi^{i(3)}$ mean
$\chi^{i34}$, $\chi^{i56}$ and $\chi^{i78}$ respectively.

  After making the necessary compensating transformation to restore the
symmetric gauge choice for the scalar fields, we find that the
scalar supersymmetry transformations become
\bea
(\delta\phi_1-\im \sinh\phi_1\delta\sigma_1)&=&
 2\sqrt{2}e^{\im\sigma_1}\bar\epsilon^{i}\chi^{j(1)}\, \epsilon_{ij}\nn\\
(\delta\phi_2-\im\sinh\phi_2\delta\sigma_2)&=
 &2\sqrt{2}e^{\im\sigma_2}\bar\epsilon^{i}\chi^{j(2)}\, \epsilon_{ij}\nn\\
(\delta\phi_3-\im\sinh\phi_3\delta\sigma_3)&=
&2\sqrt{2}e^{\im \sigma_3}\bar\epsilon^{i}\chi^{j(3)}\, \epsilon_{ij}\,.
\eea

\end{document}